\documentclass{aa}  

\usepackage{graphicx}
\usepackage{txfonts}
\usepackage{mathtools}
\usepackage{enumitem}
\usepackage{tikz,hyperref}

\definecolor{lime}{HTML}{A6CE39}
\DeclareRobustCommand{\orcidicon}{%
    \begin{tikzpicture}
    \draw[lime, fill=lime] (0,0) 
    circle [radius=0.16] 
    node[white] {{\fontfamily{qag}\selectfont \tiny ID}};
    \draw[white, fill=white] (-0.0625,0.095) 
    circle [radius=0.007];
    \end{tikzpicture}
    \hspace{-2mm}
}

\newcommand{\orcidKeerthana}{\href{https://orcid.org/0009-0001-9879-2119}{\orcidicon}}
\newcommand{\orcidEvelyn}{\href{https://orcid.org/0000-0002-2368-6469}{\orcidicon}}
\newcommand{\orcidBoris}{\href{https://orcid.org/0000-0002-1857-2088}{\orcidicon}}
\newcommand{\orcidKalina}{\href{https://orcid.org/0000-0001-5294-8002}{\orcidicon}}

\begin{document}

   \title{BUDDI-MaNGA III: The mass-assembly histories of bulges and discs of spiral galaxies}

   \author{Keerthana Jegatheesan
          \inst{1}\fnmsep\thanks{\email{ keerthana.jegatheesan@mail.udp.cl}}\orcidKeerthana
          \and
          Evelyn J. Johnston\inst{1} \orcidEvelyn
          \and
          Boris Häu\ss ler\inst{2} \orcidBoris
          \and 
          Kalina~V.~Nedkova\inst{3} \orcidKalina
          }

     \institute{
        Instituto de Estudios Astrof\'isicos, Facultad de Ingenier\'ia y Ciencias, Universidad Diego Portales, Av. Ej\'ercito Libertador 441, Santiago, Chile  
        \and 
        European Southern Observatory, Alonso de Cordova 3107, Casilla 19001, Santiago, Chile 
        \and 
        Department of Physics and Astronomy, Johns Hopkins University, Baltimore, MD 21218, USA 
        }

   \date{Received ---; accepted ---}

  \abstract
{
The many unique properties of galaxies are shaped by physical processes that affect different components of the galaxy -- such as the bulges and discs -- in different ways, and they leave characteristic imprints on the light and spectra of these components. Disentangling their spectra can reveal vital clues that can be traced back in time to understand how galaxies, and their components, form and evolve throughout their lifetimes. With BUDDI, we have decomposed the integral field unit (IFU) datacubes in SDSS-MaNGA DR17 into a S\'ersic bulge component and an exponential disc component and extracted their clean bulge and disc spectra. BUDDI-MaNGA is the first and largest statistical sample of such decomposed spectra of 1452 galaxies covering morphologies from ellipticals to late-type spirals. We derived stellar masses of the individual components with spectral energy distribution (SED) fitting using \textsc{BAGPIPES} and estimated their mean mass-weighted stellar metallicities and stellar ages using \textsc{pPXF}. With this information in place, we reconstructed the mass assembly histories of the bulges and discs of the 968 spiral galaxies (Sa-Sm types) in this sample to look for systematic trends with respect to stellar mass and morphology. Our results show a clear downsizing effect especially in the bulges, with more massive components assembling earlier and faster than the less massive ones. Additionally, in regards to comparing the stellar populations of the bulges and discs in these galaxies, we find that a majority of the bulges host more metal-rich and older stars than their disc counterparts. Nevertheless, we also find that there exists a non-negligible fraction of the spiral galaxy population in our sample with bulges that are younger and more metal-rich than their discs. We interpret these results, taking into account how their formation histories and current stellar populations depend on stellar mass and morphology.  }


   \maketitle
%

\section{Introduction}
\label{sec:intro}

The morphological classification scheme proposed in \citet{hubble1926} has brought about a search for the formation and evolution of the diverse galaxy types at various redshifts. 
Since then, it has become clear that each galaxy type is complex and diverse, built of elaborate components and sub-structures. The simplest picture and also the most standard one in the literature is that most galaxies have two structural and stellar components -- the bulge and the disc. Disc galaxies may also contain bars and spiral arms that potentially drive evolution along a particular pathway. The primary disc component is often structurally observed as an extended component with an exponential surface brightness profile \citep[for example,][]{Freeman1970ApJ, Kormendy1977ApJ}, and it is typically a region of active star formation where new stars are born from clouds of gas and dust. The bulge on the other hand is believed to be a central compact component embedded in the midst of the disc that follows a de Vaucouleurs profile \citep{devaucouleurs1948}, or more generally a S\'ersic profile \citep{sersic1968}. Traditionally, this component has been linked to old quiescent stars with high metallicities, mirroring those in elliptical galaxies, but this simple picture has been contested over the last few decades. 

Amongst these disc galaxies in the local Universe, the spiral galaxies form the predominant population, possessing a wide range of complexity in their structures. Despite their substantial contribution, the mechanisms that have shaped their components and sub-structures and the drivers of their evolution remain somewhat unresolved. Over the years, several theories have been proposed to explain their observed properties. Studying the structural and spatially resolved spectroscopic properties of these galaxies and their individual components, and tracing their star formation histories (SFH) across cosmic time, allows us to better constrain the processes involved in the formation and evolution of the components of spiral galaxies.

The classical picture follows within the hierarchical mass assembly scenario, wherein the stellar disc of a spiral forms by the collapse of gas in a rotating dark matter halo through angular momentum conservation \citep{fall&efstathiou80}. Bulges, on the other hand, tell a different story in terms of their formation pathway. Classical bulges are thought to have formed as a result of violent processes: the dissipative collapse of protogalaxies \citep{larson76}, major merger events or a series of minor mergers that rapidly exhaust the gas in star formation \citep{hopkins2009}, or by coalescence of giant gas clumps found in high redshift discs \citep{elmegreen2008, kormendy2016}. While these were the standard pictures for bulge formation for a long time, it has been challenged due to observations revealing a discrepancy in the form of {disc-like} bulges, where `bulges' simply refer to a more central and concentrated component embedded within disc galaxies. These have been traced back to gentler secular processes, incited through bar and spiral arm instabilities \citep{fisher&drory2008, kormendy2016}. 

There have been two major pathways of building galaxies described in the literature: the inside-out formation mode, whereby galaxies build up their centres first, and the outside-in formation mode, where the outermost regions form earlier and faster compared to the central parts. The information encoded in the observations of stellar spectra in a galaxy can be used to excavate physical properties of their stellar populations and ionised gas and kinematics, and this enables the study of their origins by tracing back their star formation histories. The `fossil-record' method recovers the SFH of a galaxy by realising the best combination of evolved single stellar populations (SSPs) that are constructed from observed stellar spectra, which fit the observed galaxy spectrum. This method has been adopted in recent integral-field spectroscopic (IFS) studies such as the one carried out by \citet{ibarra-medel2016}. They find an average stellar mass formation time that decreases with radius, with the central regions having built up their masses earlier than the outskirts, which supports the inside-out formation history of galaxies. This is further validated by their finding of a negative mean stellar age gradient with older stellar populations in the centre, and younger stellar populations in the outer regions, consistent with the picture of inside-out formation. Similar negative age gradients have also been found in \citet{sanchez-blazquez2014}, \citet{li2015}, \citet{goddard2017}, \citet{dominguez-sanchez2020}, and \citet{breda2020a}. More recently, \citet{lah2023} confirmed that bulge-dominated regions of galaxies are older and more metal-rich than disc-dominated regions, again supporting an inside-out formation mode.

The evolution of a galaxy across its components is expected to follow one of two main scenarios: the {inside-out} and {outside-in} quenching modes. The inside-out mechanism suggests that a quenching phase begins in the central regions of the galaxy, which then spreads outwards slowly, leading to the quenching of the disc. Furthermore, the flow of gas towards the bulge can not only fuel central star formation, but also an active galactic nucleus (AGN). Massive bulges in spiral galaxies have shown signatures of negative AGN feedback that prevents gas cooling and suppresses star formation in the centre, while the disc remains actively star forming. The disc eventually quenches due to internal secular processes unless a constant supply of gas keeps up on-going star formation. An outside-in mechanism however has also been suggested in recent studies, where the disc of the galaxy begins to halt its star formation first, while the bulge is still forming stars. Environmental effects have been determined to be the major instigators for this phenomenon: events such as ram-pressure stripping \citep{gunn&gott1972} and harassment \citep{moore1996} tend to remove the gas in the outer regions of the galaxy, where they are less strongly gravitationally bound, as they fall into a cluster or a group, quenching the outer discs first and ultimately quenching inwards to the centre. 

The extensive photometric surveys such as the Cosmic Evolution Survey  \citep[COSMOS;][]{scoville2007cosmos}, the Sloan Digital Sky Survey \citep[SDSS;][]{york2000sloan}, Galaxy and Mass Assembly \citep[GAMA;][]{driver2011gama, liske2015gama, baldry2018gama}, the Great Observatories Origins Deep Survey \citep[GOODS;][]{dickinson2003goods, giavalisco2004goods}, and the Cosmic Assembly Near-infrared Deep Extragalactic Legacy Survey \citep[CANDELS;][]{grogin2011candels, koekemoer2011candels} have allowed independent studies of major galaxy components through their decomposition into bulges and discs. With the multiple wavebands in photometric surveys, this method can be extended to fit the galaxy bulge and disc in each band, and extract their magnitudes and structural parameters. One of the tools that can do this is \textsc{GalfitM} \citep{haeussler2013megamorph, vika2014megamorph, haeussler2022galfitm}. \textsc{GalfitM} is a modified version of \textsc{Galfit} \citep{peng2002galfit, peng2011galfit}, which allows simultaneous multi-band fitting of galaxy light profiles, thereby making use of all available information at all observed wavelengths pertaining to a galaxy to fit every image. It does this by using Chebyshev polynomials of different orders to model the variation of the structural parameters with wavelength, thus ensuring a smooth and physically sensible transition across wavelength, rather than fitting each band individually. The resulting magnitudes and consequently their photometric colours can then be used to obtain the photometric stellar populations in each component by successive spectral energy distribution (SED) fitting \citep[readers can refer to the review by][]{conroy2013}. However, it has been shown that such SED fitting can suffer from an age-dust-metallicity degeneracy \citep{worthey1999, conroy2013}: the stellar ages and metallicities coming from photometric colours are highly degenerate, with the added issue of dust reddening. Where spectroscopic confirmation cannot be ensured, statistical photometric studies of large samples have the advantage of being less sensitive to catastrophic failures and misinterpretations that can happen when examining individual objects. 

To break this degeneracy, the first step would be to introduce spectroscopic information, and several approaches have been developed over the last decade to isolate the bulge and disc light from long-slit spectra. \citet{johnston2012fornax} modelled the 1D galaxy light profile of the bulges and discs of S0 galaxies in the Fornax and Virgo clusters at every wavelength of the long-slit spectra aligned along the major axis of the galaxy.  A more complex approach was taken in \citet{silchenko2012s0}, for a study of S0 galaxies with long-slit spectroscopy using SCORPIO \citep{afanasiev2005scorpio}, on the 6-m Special Astrophysical Observatory telescope, where the radius of the disc-dominated region was determined using isophotal fitting; this was followed by building a 2D model image of the disc first by masking out the inner regions, and then subtracting this from the full galaxy image to obtain the bulge parameters. Consequently these final bulge and disc parameters from this multi-step decomposition were used to extract the bulge and disc spectra from the long-slit spectrum of each galaxy. Alternatively, \citet{coccato2011kinematic} and \citet{johnston2013kinematic} adopted a kinematic decomposition of the bulge and disc, where the bulge was assumed to be dispersion supported and the disc to be rotationally supported. The drawback of long-slit spectroscopy is that the spatial distribution is not well resolved, and the bulge-disc decomposition can only occur along the major axis of the galaxy, and it would therefore lose information on non-axisymmetric features where there might be regions of renewed star formation. Therefore, any results about the stellar populations in these components are not necessarily representative of the galaxy properties as a whole. The best way to resolve this is to bring in both spatial and spectroscopic information simultaneously. With the advent of wide-field IFS such as the Multi-Unit Spectroscopic Explorer \citep[MUSE;][]{bacon2010muse} and surveys such as the Mapping Nearby Galaxies at APO \citep[MANGA;][]{bundy2015manga}, the Calar Alto Legacy Integral Field Area Survey \citep[CALIFA;][]{sanchez2012califa}, and the Sydney-Australian-Astronomical-Observatory Multi-object Integral-Field Spectrograph Galaxy Survey \citep[SAMI;][]{croom2021sami} over the past 10 years, spatially resolved spectroscopic bulge-disc decomposition has now become a possibility.

In the last few years, several studies have already successfully implemented the use of integral field units (IFUs) to extract stellar populations of different regimes in the galaxy. For instance, \cite{fraser2018manga} used the photometric decomposition of imaging data to identify the bulge and disc-dominated regions in S0 galaxies in the MaNGA survey, and \cite{barsanti2021sami} used a similar technique on S0 galaxies in the SAMI survey to measure their stellar populations. 

An unfortunate drawback to these techniques is the fact that there exists a low level of contamination or overlap between the components, which can ultimately bias the stellar ages to be artificially younger or older when measured. In order to reduce this contamination, there have been several studies (which we discuss in detail below) that extend the idea of multi-waveband light profile fitting to the many wavelengths of IFU datacubes to cleanly isolate the light and the spectra of each component.

The first code to accomplish this is BUlge-Disc Decomposition
of IFU data \citep[BUDDI;][]{johnston2017buddi}, which creates wavelength-dependent models of each component by fitting the galaxy in each image slice of the IFU datacube. For a brief overview of \textsc{BUDDI}, readers can refer to Sect. \ref{sec:galaxy-decomp} of this paper and for more details on the fitting process, readers can refer to \cite{johnston2022buddi1}. This allows a cleaner extraction of the spectra of the components with minimal contamination from either component on the other. BUDDI was first tested successfully on eight S0 galaxies in MUSE \citep{johnston2021muse}, where the decomposition was made for the bulge, disc, and lens components. Another code that works along the same lines is C2D \citep{mendez-abreu2019c2d}, which has also been successfully tested with a bulge and disc decomposition on CALIFA galaxies \citep{mendez-abreu2019discs, mendez-abreu2021morph}. 

Following these successful tests using improved techniques, the next step would be to extend the study of galaxy components to a more statistical sample to cover a better range of morphologies and stellar masses. The present study forms the third paper within the BUDDI-MaNGA framework. It mainly relies on the fits and catalogue built from a sub-sample of MaNGA galaxies in SDSS DR17, which decomposes the spectra of the galaxies into their bulge and disc components, along with the publicly available value-added catalogues (which provide additional information on the physical properties of the MaNGA galaxies based on different studies), and corresponding SDSS optical imaging data products that provide a range of ancillary information. This final sample of fits forms by far the largest statistical sample of IFU-based decomposition with 1452 galaxies containing bulge and disc spectra. The technical details of the methodology can be found in \citet{johnston2022buddi1}, which describes the technique applied to SDSS-MaNGA DR15. A brief overview is provided in Sect.~\ref{subsec:buddi_decomp} along with the modifications implemented in the new version. In this paper, we turn our focus to reconstructing the stellar mass assembly histories of bulges and discs in spiral galaxies, to understand the pathways through which they have built up their mass over cosmic times. Moreover, by comparing the stellar populations hosted in the bulges and discs of these galaxies, we explore the physical processes that have driven their evolution. 

This paper is organised as follows. Sect.~\ref{sec:data} outlines the data and catalogues used in this study, Sect.~\ref{sec:galaxy-decomp} gives a brief overview of the bulge-disc decomposition technique with \textsc{BUDDI}, along with the selection criteria for the final BUDDI-MaNGA DR17 sample. Sect.~\ref{sec:methods} describes the methodology we employed to estimate the stellar masses of bulges and discs, and the full-spectral fitting procedure of the resulting decomposed spectra to derive physical parameters of stellar populations. The main results of this work are presented in Sections \ref{sec:galaxy_mah} and \ref{sec:stellar_pops}, describing the trends observed in the mass assembly histories of both components as a function of different physical properties and the analysis of their stellar populations. Section \ref{sec:discussion} discusses the results in the context of previous related studies, and we draw our conclusions in Section \ref{sec:conclusions}. Throughout this work, unless stated otherwise, all magnitudes are relative to the AB system \citep{oke&gunn1983}, and we have adopted the flat $\Lambda$CDM cosmology with $H_0 = 70 \; \mathrm{kms^{-1}Mpc^{-1}, \; \Omega_m = 0.3, \; \Omega_{\Lambda} = 0.7}$. 

\section{Data and catalogues}
\label{sec:data}

In this section, we list the datasets and catalogues that we use throughout this paper.   

\subsection{SDSS}
\label{subsec:SDSS_survey}

SDSS is an imaging and spectroscopic redshift survey, that employs the dedicated 2.5m Sloan Foundation telescope \citep{gunn2006sdss} at Apache Point Observatory (APO) in New Mexico. SDSS has been an ongoing effort for over two decades to provide optical imaging for approximately one quarter of the sky in the Northern Galactic Cap. It provides deep photometry ($r<22.5$) in five optical passbands: $ugriz$ \citep{fukugita1996sdss, smith2002sdss}. In this work, we use the imaging data products from the SDSS Data Release 17 \citep{abdurrouf2022sdss17}.

\subsection{The MaNGA IFU survey}
\label{subsec:manga_survey}

This work centres on MaNGA \citep[Mapping Nearby Galaxies at APO;][]{bundy2015manga} data released in the SDSS Data Release 17. MaNGA is an integral field spectroscopic survey, which is a part of the SDSS-IV surveys \citep{blanton2017sdss}. Using a modification of the BOSS spectrograph \citep{smee2013boss}, the fibres are bundled into hexagons \citep{drory2015ifu}  marking different IFU sizes to cover the galaxies, ranging from 19 fibres to 127 fibres, with 12\arcsec and 32\arcsec diameters respectively. The final public data release holds a total of 10\,127 galaxies within the redshift range $0.01 < z < 0.15$ \citep{yan2016mangaifs}, and a flat mass distribution within $10^9 - 10^{11}M_{\odot}$. The IFU size for each galaxy
is selected such that it covers the galaxy out to $\sim1.5R_e$ for $\sim66$ per cent
of the total galaxy sample, and out to $\sim2.5R_e$ for the rest. The spectra have a continuous wavelength coverage spanning from 3\,600~$\AA$ to 10\,300~$\AA$, and a spectral resolution of $R \sim 1\,400$ at 4\,000~$\AA$ to $R \sim 2\,600$ at 9\,000~$\AA$ \citep{drory2015ifu}. 

\begin{figure*} 
    \includegraphics[width=0.9\textwidth]{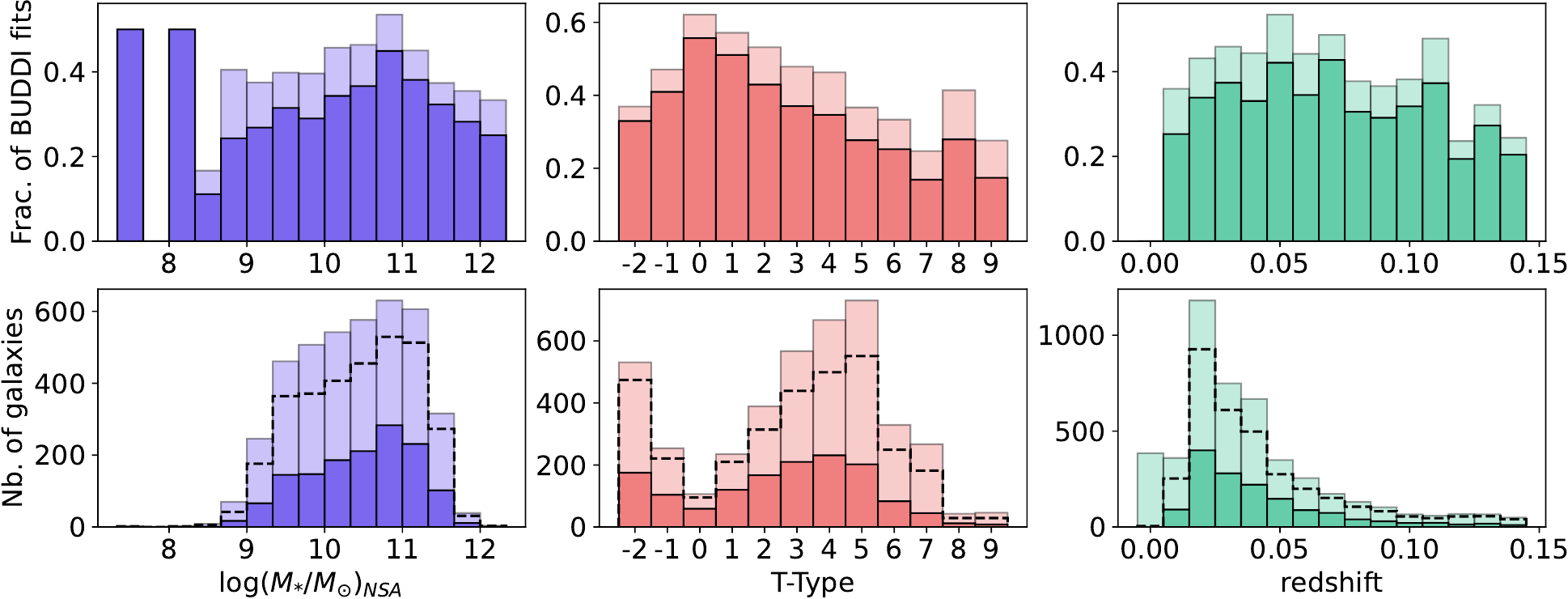} 
    
    \caption{Histograms depicting the distributions of physical properties of different samples used in this study. Lower panels: Distributions of the galaxy stellar mass, morphological T-Type, and redshift in the successful fits in the new BUDDI-MaNGA DR17 sample, as defined in Sect \ref{subsec:sample_selection} (darker shades), and for all the MaNGA galaxies observed with the 91 and 127 fibre IFUs (lighter shades). The dashed outlines depict the distributions of the galaxies that were successfully fit with \textsc{PyMorph} in the MPP-VAC-DR17, which determined the initial set of objects from which we built the BUDDI-MaNGA sample. The T-Types of the ellipticals and S0s from the VAC are reassigned indices of -2 and -1, respectively, for continuity in the distribution. Upper panels: Distributions of the fraction of successful BUDDI fits with respect to the \textsc{PyMorph} fits sample (lighter shades), and a fraction of successful BUDDI fits with respect to the MaNGA sample with the largest IFUs (darker shades).}
    \label{fig:overview}
\end{figure*}

\subsection{Galaxy Zoo: 3D}
\label{subsec:gz3d_cat}
The \textsc{Zooniverse}\footnote{\url{https://www.zooniverse.org/lab}} crowd-sourcing platform was utilised by the MaNGA team to construct GZ:3D, a project where citizen scientists could contribute to creating spaxel maps identifying various sub-features in galaxies, such as galactic bars and spiral arms. Aside from this, they also identify and mark the presence of foreground stars and locate the central spaxels of the galaxy. These properties have been gathered into a set of masks available in the value-added catalogue, that has been described in greater depth in \citet{masters2021gz3d}, which we use to mask out foreground stars as described in Sect \ref{subsec:buddi_decomp}.

\subsection{MaNGA PyMorph Value-Added Catalogue}
\label{subsec:pymorph}
The MaNGA PyMorph Value-Added Catalogue DR17 \citep[MPP-VAC-DR17;][]{dominguez-sanchez2022pymorph} is a table that provides photometric structural parameters for galaxies in SDSS-MaNGA DR17. These estimates come from modelling the 2D surface profiles of the SDSS galaxies using the \textsc{PyMorph} \citep{vikram2010pymorph, meert2013decomp, meert2015decomp, meert2016decomp} fitting software, which employs \textsc{SExtractor} \citep{bertin&arnouts1996sextractor} and \textsc{Galfit} \citep{peng2002galfit, peng2011galfit}. This VAC provides separate estimates for the $g,r,i$ bands of the SDSS images of 10\,127 galaxies. The structural parameters were measured for both single S\'ersic fits and S\'ersic + Exponential (two components) fits. The catalogue additionally provides a flag where either or both fits failed, and in the case of galaxies that have successful fits with both models, denotes the best model preferred amongst them. This version of the VAC has been updated from the MPP-VAC-DR15 \citep{fischer2019pymorph}, which was built on SDSS DR15, and \citet{fischer2019pymorph} details the exact methodology and algorithm used in the \textsc{PyMorph} fits. We use these values as starting parameters for the BUDDI-MaNGA fits in an automated fashion.

\subsection{MaNGA Visual Morphologies from SDSS and DESI images}
\label{subsec:morph_cat}

This is a  value-added catalogue provided with the MaNGA data products that provides visual morphological classification of the MaNGA DR17 galaxies based on the $r$-band images from SDSS and the Dark Energy Spectroscopy Instrument (DESI) Legacy Survey \citep{dey2019}, using a new digital re-processing technique described in \citet{vazquez-mata2022mangamorph}. While the two widely used morphology catalogues are the Galaxy Zoo visual classifications \citep{masters2020galaxyzoo} and the deep learning classifications in \citet{dominguez-sanchez2018dl, dominguez-sanchez2022pymorph}, the former has a significant fraction of galaxies that have no definitive classification, and the latter can be skewed by image quality. This VAC provides both Hubble types and the T-Type indices associated to each of them, along with bar identification. In this work, we use both indices simultaneously, but our interpretations are often based on the Hubble types. 

\subsection{The Nasa Sloan Atlas (NSA) catalogue}
\label{subsec:nsa_cat}

The NSA is a catalogue containing useful derived quantities including galaxy stellar mass, redshift, magnitudes and line flux measurements. The catalogue was built from SDSS DR8 photometry, combined with UV photometry from GALEX \citep[Galaxy Evolution Explorer;][]{blanton2007galex}. We use the NSA v1 catalogue, which extends out to $z=0.15$, and covers the redshift range of the galaxies in MaNGA.

\section{Galaxy decomposition in SDSS-MaNGA DR17}
\label{sec:galaxy-decomp}

\subsection{Input sample selection and data preparation}
\label{subsec:buddi_sample}

The input sample to our analysis is taken from the SDSS-MaNGA DR17, which consists of 11 273 galaxy datacubes in total, out of which 10 010 are unique galaxies with high-quality observations. From this, we select only those galaxies having the two largest IFU sizes with 91 and 127 fibres, since \citet{johnston2017buddi} find that two-component fits to the galaxies observed with smaller IFU sizes (19, 37 and 61 fibres) become unreliable owing to their small fields of view and availability of spaxels. This leaves us with 4661 galaxies that can potentially be fit with either a single S\'ersic, a S\'ersic + Exponential model, or both.

From this sample, we further narrowed down our selection to the galaxies that were flagged to have successful fits with these models in the MPP-VAC-DR17 (readers can refer to Sect. \ref{subsec:pymorph}). We also eliminate galaxies with non-physical fit parameters, where the $R_e$ (effective radius) is less than 1 pixel in this catalogue, which results in 4051 objects. These parameters from the MPP-VAC-DR17 would serve as initial values for the \textsc{GalfitM} fits in \textsc{BUDDI} (readers can refer to Sect. \ref{subsec:buddi_decomp}). 

For the fits to the datacubes, we further require a bad pixel mask, in order to mask any pixels in the rectangular fits files that were not covered by the hexagonal IFU bundles. This is provided as an extension in the MaNGA datacubes in their data releases. Additionally, this extension also includes sufficient masks of foreground stars, which were created by visual inspection by the MaNGA team and complemented by GZ:3D (readers can refer to Sect. \ref{subsec:gz3d_cat} and \ref{subsec:star_mask_effect}). Using this information in the fitting procedure is the significant modification for improving the fits from the previous run on DR15 galaxies (readers can refer to Sect.~\ref{sec:discussion}).

To improve the fits, a point spread function (PSF) datacube was created for each galaxy to be convolved with the fit to each image in the datacube. The (reconstructed) PSF for the $griz$ bands are also provided in the datacube, from which the PSF at each wavelength was created through interpolation. \textsc{GalfitM} and by extension, \textsc{BUDDI}, allows a sigma datacube to be provided in order to estimate the flux uncertainty in each pixel more accurately. Lastly, the MaNGA datacubes contain an inverse variance datacube (IVAR) from which we derive the sigma datacube as $\mathrm{1/\sqrt{IVAR}}$.

\subsection{Bulge-disc decomposition with BUDDI}
\label{subsec:buddi_decomp}

The decomposition of the MaNGA galaxies into their bulge and disc components itself is done using \textsc{BUDDI}, an \textsc{IDL} wrapper for \textsc{GalfitM} that can handle the many wavelengths of IFUs and uses all available data simultaneously in order to optimise this task.

As described in \citet{johnston2017buddi} and \citet{johnston2022buddi1}, \textsc{BUDDI} starts by measuring and normalising the galaxy kinematics from the IFU datacubes. This step is important since \textsc{GalfitM} can only model symmetric structures, and failure to normalise the kinematics in this way can lead to artefacts in the final spectra. The Voronoi binning technique described in \citet{cappellari&copin2003} was used to bin the datacubes for this purpose. The kinematics of the binned spectra were measured with \textsc{pPXF}, making use of the Medium
resolution INT Library of Empirical Spectra (MILES) stellar library \citep{sanchez-blazquez2006miles} to determine the line-of-sight velocities and velocity dispersions. Following this, the spectrum in each spaxel was shifted to match the line-of-sight velocity of the galaxy centre, and broadened to match the maximum velocity dispersion in the galaxy. This way, the IFU cube is ideally set up for BUDDI. 

The code then follows a three-step process to derive the spectra of each component: 
\begin{itemize}
\item In the first step, a single broad-band image of the galaxy is created by stacking the entire datacube, and modelled with \textsc{GalfitM} with both a single S\'ersic profile to model the galaxy as a whole, and a S\'ersic + Exponential profile fitting the bulge and disc, respectively. The $r$-band structural parameters from the MPP-VAC-DR17 \citep{dominguez-sanchez2022pymorph} were used as the initial parameters in \textsc{BUDDI}, which also helped contain the fits to a reasonable sample eliminating irregular morphologies with no distinct bulges and discs. The MPP-VAC provides flags where either the single-component model (S\'ersic) or the two-component model (S\'ersic + Exponential) or both have failed (readers can refer to Sect. \ref{subsec:pymorph}), which further constrains our sample to those galaxies whose surface brightness profiles can be sufficiently modelled. 
\item To better refine the fit parameters as a function of wavelength, specifically to derive the Chebyshev polynomials used in \textsc{GalfitM}, the datacube was rebinned in terms of wavelength into a series of ten narrow-band images, following the approach outlined in \cite{johnston2022buddi1}. These ten images were again modelled with \textsc{GalfitM} by fixing the initial parameters to those derived in the previous step. Additionally, this step is where the variations due to wavelength are introduced through the above-mentioned Chebyshev polynomials. A polynomial of order 1 was selected for the parameters $R_e$, $n_{\textit{S\'ersic}}$, $q$, and $PA$, which forces them to be constant with wavelength. This is because \citet{haeussler2022galfitm} find that the colour differences that exist within galaxies are small compared to differences between the components, which makes accurate measurements difficult; therefore, using a first order polynomial to model the structural parameters allows us to derive the spectra of the bulge and disc components more easily. 
\item Finally, in a third and last step, BUDDI fixes all the structural parameters to those derived in the previous step, and by leaving only the magnitudes free, derives the component magnitudes in each individual image slice of the datacube. This allows us to cleanly extract the 1D spectra of each component.

\end{itemize}

\subsection{Selection of final sample}
\label{subsec:sample_selection}

With BUDDI, we obtained a sample of 2699 galaxies that were successfully fit with the SE model (67\% of the parent sample with reasonable \textsc{PyMorph} parameters), that is, the fit converged on a final solution and did not crash. To refine this sample, we used the selection criteria outlined in \citet{johnston2022buddi1, johnston2022buddi2}, which led to a final sample of fits with acceptable structural parameters recovered in the BUDDI fits:

\begin{enumerate}[label=(\roman*)]
    \item $\Delta mag_{r} \; \leq 2.5$ 
    \item $0.25\arcsec \; \leq \; R_e \; \leq 50\arcsec$
    \item $0.205 \; \leq n_{\textit{S\'ersic}} \; \leq 7.95$
    \item $0.1 \; \leq q \; \leq 1$,
\end{enumerate}

\noindent where $\Delta mag_{r}$ is defined as the difference in the magnitudes over the $r$-band wavelength range between the S\'ersic and exponential components, which corresponds to the fainter component accounting for 10\% of the total light. $R_e$ is the half-light radius, $n_{\textit{S\'ersic}}$ is the S\'ersic index of the S\'ersic component, and $q$ is the axis ratio of both components. This selection criteria led to a sample of 1452 galaxies with good fits and physically reasonable fit parameters, for which the bulge and disc spectra have been extracted. Although single S\'ersic (one-component) fits were also performed, they are not relevant to this particular study and are not discussed here. More details on the methodology, choice of selection criteria, and derived spectra can be found in \citet{johnston2022buddi1}. Furthermore, Sect. 4.1 in \citet{johnston2022buddi1} details the common reasons why fits can fail with BUDDI.

Figure~\ref{fig:overview}  shows the general overview of the BUDDI-MaNGA DR17 sample, with the distributions of galaxy stellar mass, T-Type, and redshift (darker shades in the histograms). The distributions for all the MaNGA galaxies that were observed with the 91 and 127 fibre IFUs are shown as the lighter histograms. From these, the galaxies that were successfully fit with a S\'ersic + Exponential model with \textsc{PyMorph} in the MPP-VAC-DR17 are shown by the dashed black outlines. The upper panels show the corresponding fraction of successful BUDDI fits with respect to the \textsc{PyMorph} sample (lighter shades), and to the MaNGA sample (darker shades). These fractions allow us to easily identify any biases that may exist purely in our sample. The galaxy stellar masses and redshifts were taken from the NSA catalogue (readers can refer to Sect.~\ref{subsec:nsa_cat}), the T-Types from visual morphology value-added catalogue described in Sect.~\ref{subsec:morph_cat}. The sample consists predominantly of galaxies within $2 \leq$ T-Type $\leq 6$, which correspond to Sab - Sc type galaxies, with a peak at T-Type = 4 (Sbc, late-type spiral). The galaxy stellar masses of the sample show hints of a bimodality in the distribution, with a maximum peak in the high mass end at $M_{*} \sim 10^{11} M_{\odot}$, and a smaller peak in the lower mass end at $M_{*} \sim 10^{9.8} M_{\odot}$. The BUDDI-MaNGA sample closely traces the distributions of the \textsc{PyMorph} sample and the MaNGA sample. Moreover, the upper panels show no significant biases that are specific to the BUDDI-MaNGA sample that are not already due to the selection effect in the MaNGA and \textsc{PyMorph} samples.

\subsection{Caveats}
We acknowledge a few caveats in the spectro-photometric modelling which are related to the assumptions we have made.

\begin{itemize}
    \item For one, we assume that the surface brightness profile of the disc is always purely exponential from the outskirts all the way to the centre of the galaxy. However, recent studies debate the validity of this premise. \citet{breda2020b} found that a significant fraction of the discs in their late-type galaxy sample in fact show a down-bending of their light profile within the radius of the bulge. Their analysis revealed that for a third of their sample, the bulge SED showed negative fluxes in the blue end of the spectrum when a standard exponential disc model was subtracted from the galaxy SED.

    \item We also make a simplistic assumption on the structural components of the galaxies to only have a well-defined bulge and disc. This might not reflect the true surface brightness profile of the complexity of components such as spiral arms and bars. For example, excluding a model for the bar can affect the structural parameters of the components, such as overestimating the bulge-to-disc ratio (B/D) and the Sérsic index of the bulge \citep{laurikainen2004}.\citet{erwin2021} find that modelling a barred galaxy with a bulge and disc overestimated the fraction of light from the bulge by a factor of 4 and 100 for their two galaxies.  While this can have an impact on the extracted spectra, it is also important to note that the low spatial resolution of MaNGA does not permit us to isolate the bars clearly and model them as a separate component. Similarly, the presence of a nuclear bar in the central region of the galaxy could cause the bulge to appear more elongated if not taken into account, which can affect the estimation of bulge structural parameters. On that note, \citet{mendez-abreu2008} investigated the influence of nuclear bars by adding them to simulated images and studying their structural parameters through photometric decomposition, to observe if the bulges showed signs of elongation. They found that the majority of the bulges were in fact circular and not significantly impacted by the presence of a nuclear bar in their midst. Therefore, we expect these results to also hold in the present study. 
    
\end{itemize}

Our assumptions for this work are primarily based on the work of \citet{fischer2019pymorph} and \citet{dominguez-sanchez2022pymorph}, which as we mentioned earlier in Sect. \ref{subsec:buddi_decomp}, served as our initial parameters for the fits. However, testing the alternate or additional premises would require this large sample of galaxies to be fit on a case-by-case basis, which would defeat the purpose of an automated fitting procedure.

\section{Methods}
\label{sec:methods}

\subsection{Estimating bulge and disc stellar masses}
\label{subsec:mass_estimation}

Due to the small field of view of the MaNGA IFUs, the light from the outskirts of the galaxies can often be lost. As a result, any mass estimates derived using the bulge and disc fluxes from \textsc{BUDDI} may be biased due to the loss of the light profile information in these outer regions. 

Therefore, in this work, the masses of the bulges and discs were estimated through SED fitting to the photometry derived from SDSS imaging (Sect. \ref{subsec: bagpipes_masses}). This data not only shows improved imaging resolution compared to the MaNGA data, but ensures broad-band photometry necessary for the SED fitting routines. For consistency, we feed to  \textsc{GalfitM} the GZ:3D foreground star masks that were created on the SDSS images, to align with the data preparation approach in \textsc{BUDDI} (readers can refer to Sect. \ref{subsec:buddi_sample}). The exact approach we employ is explained below. An alternative approach employing the relation between optical colour and the stellar mass-to-light ratio was also explored, and will be detailed below in Sect. \ref{subsec: cmr_masses}. 

The SDSS images were modelled with two components using \textsc{GalfitM} with the structural parameters fixed to those derived in the \textsc{BUDDI} fits (readers can refer to Sect.~\ref{subsec:buddi_decomp}), to derive the magnitudes in the $ugriz$ bands. These magnitudes were fit with polynomials of order 5 which were allowed complete freedom as a function of wavelength (freedom in 5 bands). By fixing the structural parameters in this way we ensure that we are measuring the masses for the same components derived from the IFU datacubes, with the caveats discussed in Sect. \ref{caveats_masses}. It must be noted that not all the galaxies in the final BUDDI-MaNGA S\'ersic + Exponential sample (from Sect.~\ref{subsec:sample_selection} had a successful fit with \textsc{GalfitM} to the SDSS imaging data. From the 1452 selected galaxies in BUDDI, we were able to recover the fits for 1390 galaxies. From this, we further selected the galaxies that had $0.1 \leq (B/T)_{r} \leq 0.9$, which corresponds to the $r$-band magnitude selection we imposed in Sect.~\ref{subsec:sample_selection}. With this in place, we have 1312 galaxies with reasonable fit parameters to both SDSS images and MaNGA datacubes, for which we can now perform SED fitting to retrieve their stellar masses. The resulting observed magnitudes were then converted into fluxes in units of $\mathrm{\mu Jy}$ using a zero-point magnitude of $23.9$, 
to be fed as input to the SED fitting code described below. \citet{haeussler2013megamorph} find that \textsc{GalfitM} underestimates the errors by a factor of $\sim2-2.5$, and we increase all magnitude errors by a factor of 3 to be conservative, as suggested in \cite{nedkova2021mzr}. This needs to be included in the SED fitting, since the codes heavily rely on the uncertainties, and a value has to be provided which is as realistic as possible.  

\begin{figure}[h!]
    \centering
    \includegraphics[width=0.99\columnwidth]{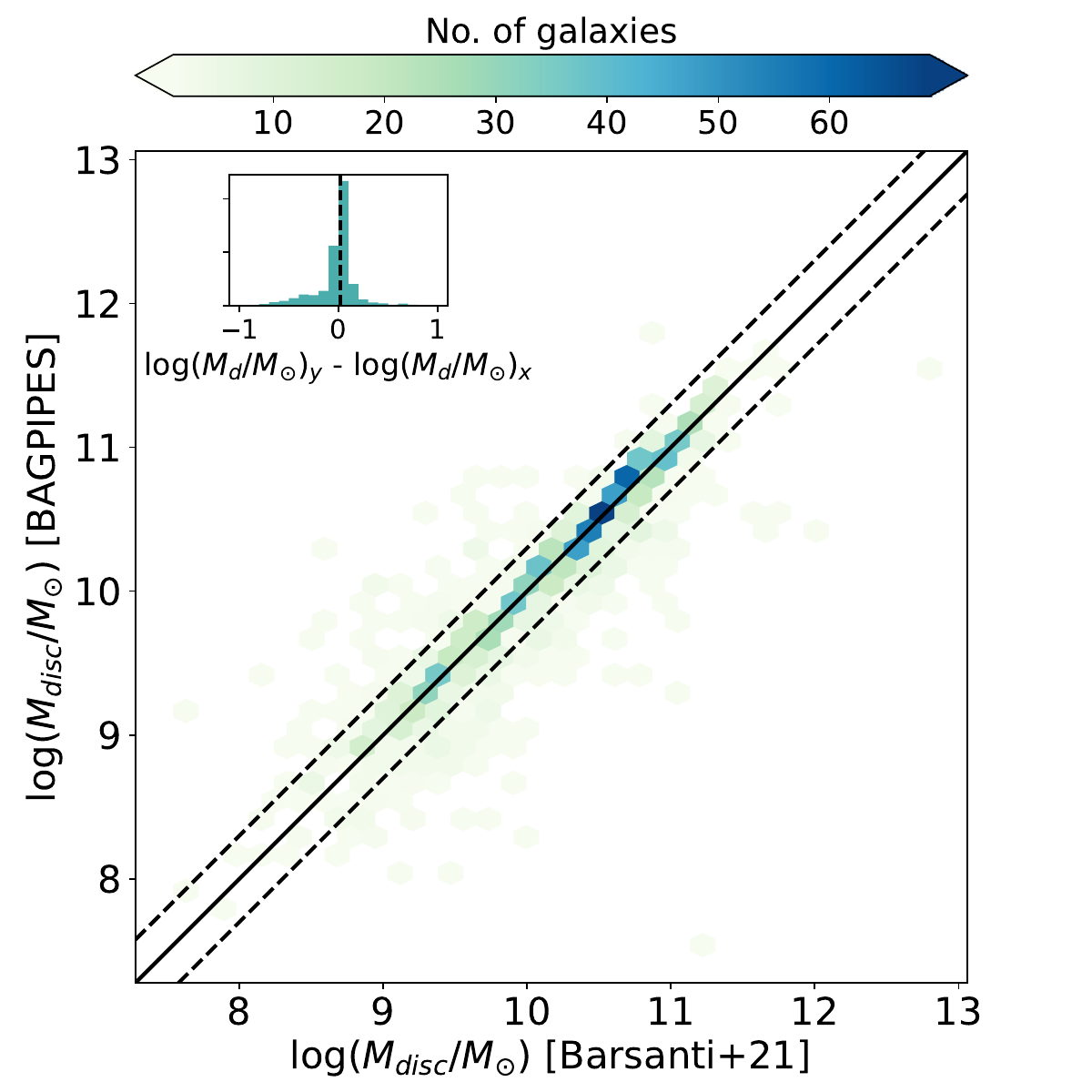}
    \caption{Density plot of the disc masses estimated by \textsc{BAGPIPES} and the colour-mass relation in \citet{barsanti2021sami}. The black diagonal line shows the 1:1 correspondence, and the dashed lines mark the $\pm0.3$ dex offset from it. The inset shows the distribution of the difference between each \textsc{BAGPIPES} estimate and the CMR estimate, with the black dashed vertical line marking the median offset.}
    \label{fig:disc-masses}
\end{figure}

\begin{figure}[h!]
    \centering
    \includegraphics[width=0.99\columnwidth]{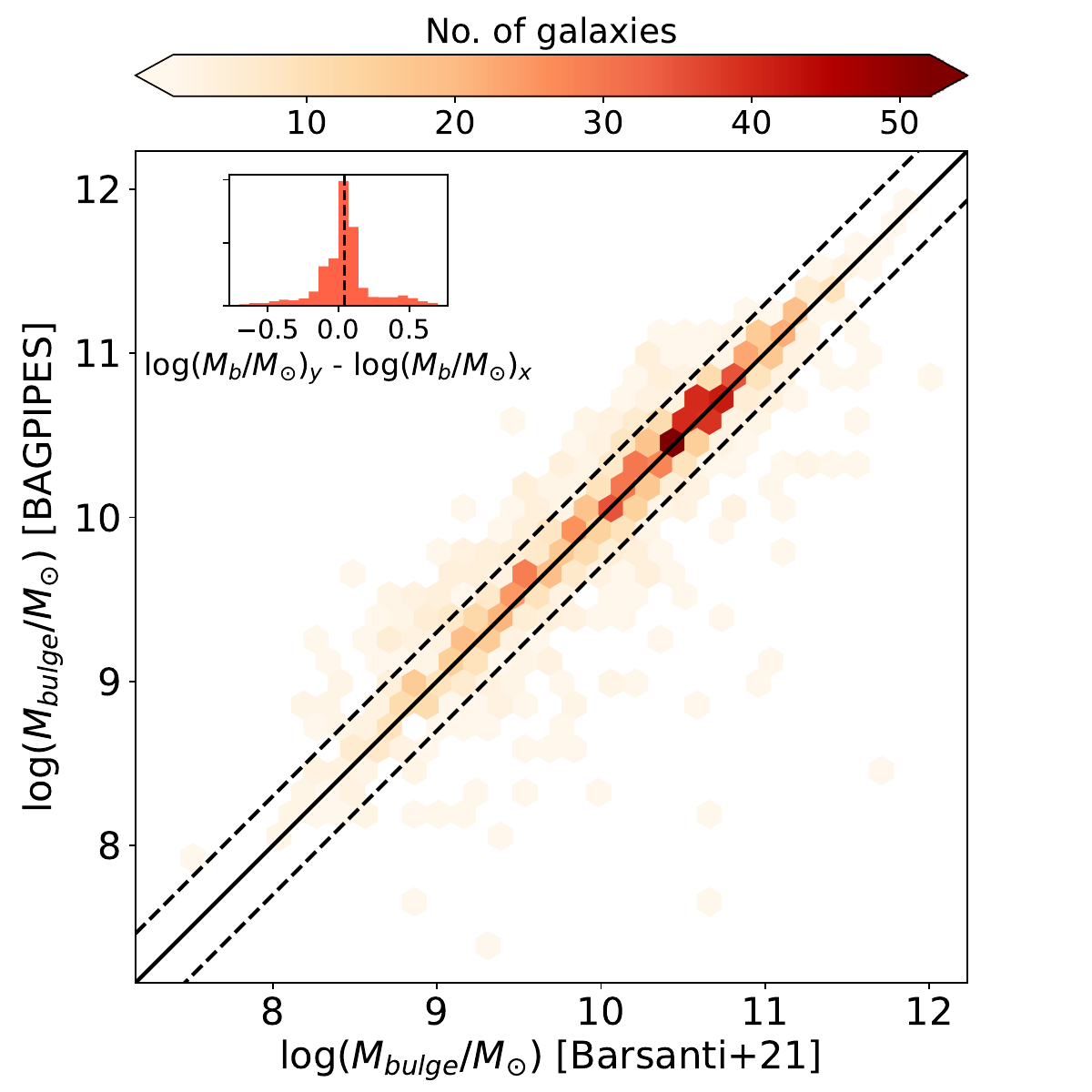}
    \caption{Similar to the density plot for Figure \ref{fig:disc-masses} but showing the bulge masses.}
    \label{fig:bulge-masses}
\end{figure}

\begin{figure}[h!]
    \centering
    \includegraphics[width=0.99\columnwidth]{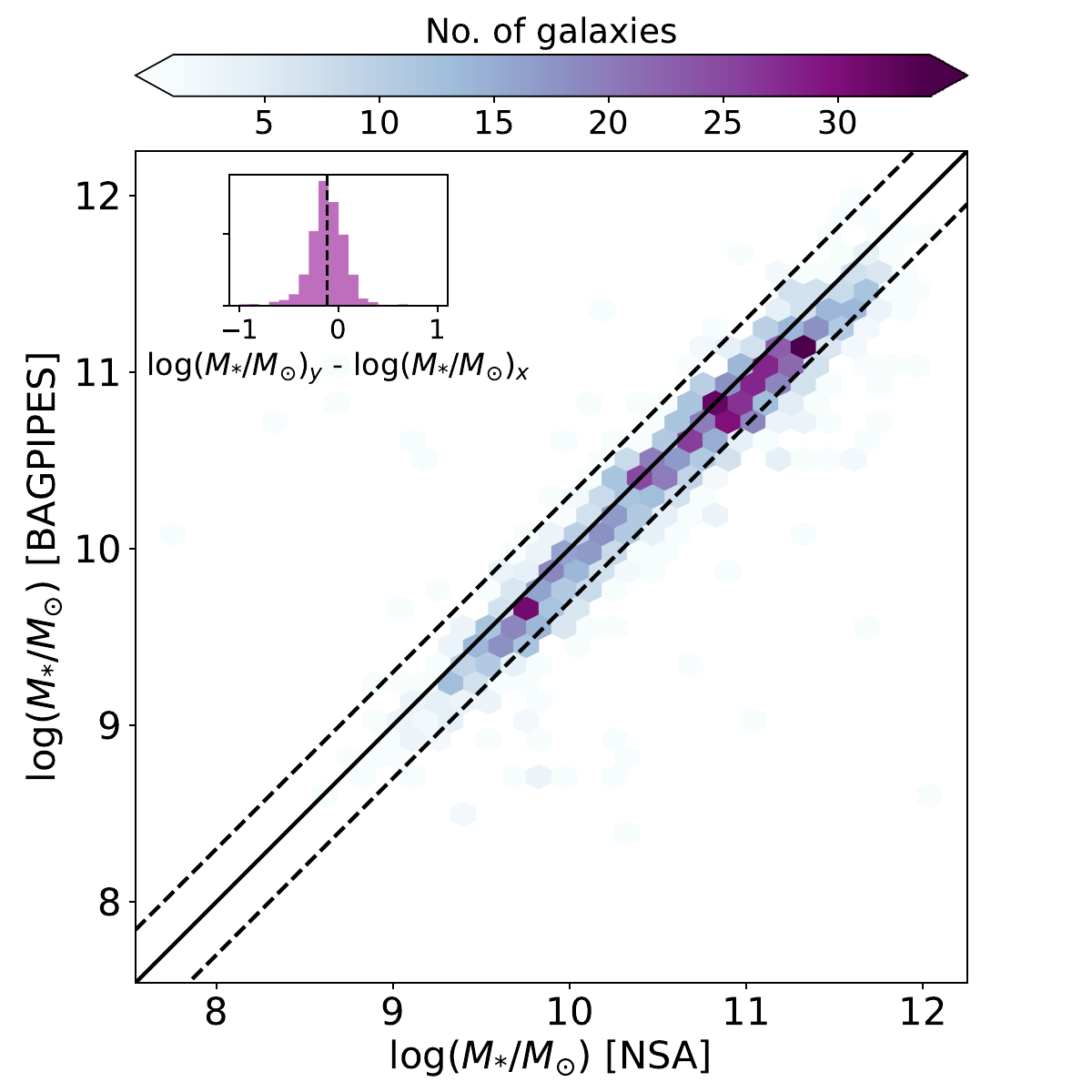}
    \caption{Similar to the density plot for Figure \ref{fig:disc-masses} but showing the total galaxy stellar masses.}
    \label{fig:total-masses}
\end{figure}

\subsubsection{\textsc{BAGPIPES} derived masses}
\label{subsec: bagpipes_masses}

The SED fitting was performed using Bayesian Analysis of Galaxies for Physical Inference and Parameter EStimation \citep[BAGPIPES;][]{carnall2018bagpipes}. Based on \textsc{Python}, \textsc{BAGPIPES} is a versatile tool that can fit both observed photometric and spectroscopic SEDs simultaneously to complex galaxy model spectra to obtain a probability distribution function (PDF) for each key physical property that can be derived with it.

We set up the code to assume the stellar population synthesis models described in \citet{bruzual&charlot2003}, built with an \citet{kroupa&boily2002imf} initial mass function (IMF). Since our main objective with \textsc{BAGPIPES} was to estimate stellar masses, we simply model the stellar continuum imposed by the photometric bands, without providing spectroscopic information pertaining to emission lines. For our redshift range, the photometric bands provide reliable stellar mass estimates because they sample the Balmer break. For the fitting, we assume an exponentially declining SFH ($\tau$ model) of the form defined in \citet{carnall2019sfh}:
\[
    \mathrm{SFR}(t)= 
\begin{dcases}
    \frac{\mathrm{exp}(-t-T_0)}{\tau},&  t>T_0\\
    0,              & t<T_0
\end{dcases}
\]
  where $T_0$ is the time at which star formation is expected to jump from its initial value of 0 to its maximum value, and $\tau$ is the e-folding timescale with which the star formation declines exponentially. \textsc{BAGPIPES} assumes a uniform prior for all parameters by default; imposing different priors may result in different estimates for the physical parameters and a detailed analysis can be found in \citet{carnall2019sfh}. We allow the stellar ages to vary between 10 Myr and the age of the Universe, 13.8 Gyr. $\tau$ is allowed to vary between 100 Myr and the age of the Universe, while a solar metallicity of 0.02 is fixed. We assume a Calzetti dust attenuation law \citep{calzetti2001dust}, allowing $A_v$ to vary between 0 and 2 mag. The prior distributions for the parameters are taken to be flat, following the default setting in \textsc{BAGPIPES}. The redshifts of the galaxies are set and fixed to the estimates from the MSR-VAC-DR17 \citep[MaNGA Spectroscopic Redshifts Value Added Catalog;][]{talbot2018specz}. For the galaxies that do not have a spectroscopic redshift from this catalogue (344 objects), the photometric redshifts from the NSA catalogue \footnote{\url{http://www.nsatlas.org}} were used instead. The Bayesian formalism integrated into \textsc{BAGPIPES} results in a posterior distribution for the stellar masses (and in principle for all physical parameters), leading to reliable uncertainty estimates. We take the median values as the parameter estimates, with the difference between the $16^{th}$ and $84^{th}$ percentiles as the uncertainties. We note that BAGPIPES fails in $\sim4\%$ of the bulges and the discs, mostly due to unfeasible errors on the magnitudes estimated by \textsc{GalfitM}, or due to a simplistic assumption of a parametric SFH for both components.

\subsubsection{Mass estimates derived from the colour-mass relation}

\label{subsec: cmr_masses}
A final comparison set was generated estimating the stellar masses purely on the basis of their optical colours and observed magnitudes in a maximum of two bands, without imposing any assumptions on the star-formation history or using stellar population synthesis models. We followed the prescription mentioned in \citet{barsanti2021sami}:
\begin{multline}
\mathrm{log_{10}}(M_*/M_\odot) = -0.4\,i + 2\,\mathrm{log_{10}}(D_L/10) - \mathrm{log_{10}}(1+z) + \\ (1.2117-0.5893\,z) + (0.7106-0.1467\,z) \times (g-i)
\end{multline} 

\noindent where $z$ is the galaxy redshift from NSA, and $D_L$ is the luminosity distance in parsec. The resulting stellar masses of the bulges and discs estimated from both methods show excellent agreement and little scatter.  
Figures~\ref{fig:disc-masses}, ~\ref{fig:bulge-masses}, ~\ref{fig:total-masses} show the comparison of these estimates (for the disc, bulge, and total mass respectively) with different methods, and it is clear the majority of them agree well with each other within 0.3 dex. The small scatter and offset can be attributed to the differences in the assumptions and the physics that go into the template fitting codes, as well as the fact that these codes make use of magnitudes in 5 bands and model the SED of the galaxies in the optical range, while the colour-mass relation only uses 2 magnitudes. We define the total galaxy mass for our sample as a simple sum of the bulge and disc stellar masses. In Figure \ref{fig:total-masses}, our estimates are compared to the total galaxy stellar masses in the NSA catalogue (converted to the cosmology parameters we assume in this paper), which were estimated from K-correction fits to S\'ersic fluxes.

For this work, we use the \textsc{BAGPIPES} estimates of stellar masses whenever we have them, and the colour-mass relation estimates for the ones where \textsc{BAGPIPES} failed to fit the galaxy SED. The sample now contains 1312 objects with physically feasible component masses. It must be noted that only the \textsc{BAGPIPES} estimates have associated errors.

\begin{figure*} 
    \includegraphics[width=0.9\textwidth]{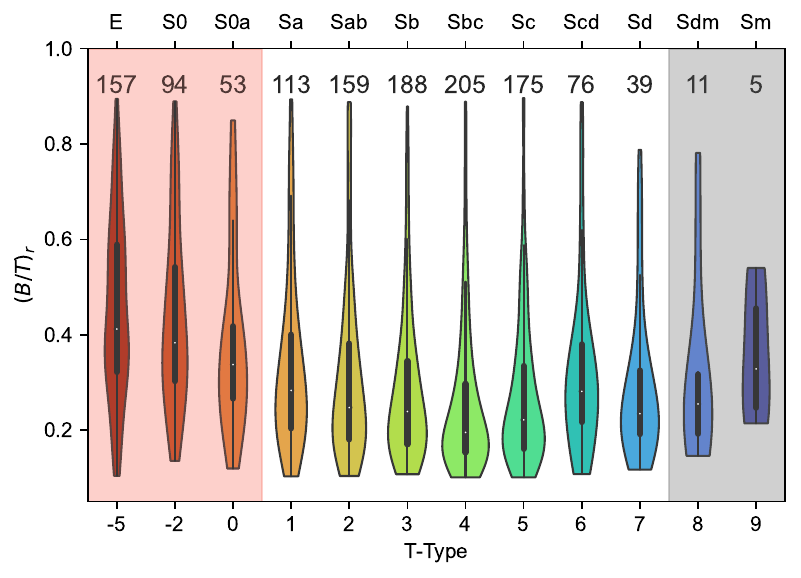} 
    
    \caption{Bulge-to-total ratio of all galaxies in the $r$-band, as a function of morphology. The lower x-axis is labelled by the T-Type index, and the upper x-axis by the corresponding Hubble types. The inner parts of the violins contain a box and whisker plot. The white circles in the violin plots represent the median bulge-to-total ratios, while the thick black bar defines the interquartile (25th - 75th percentile) range of the distribution. The limits on the thin black bar extend to 1.5 times the interquartile range. The upper and lower ends of the violins mark the highest and lowest values of $(B/T)_r$ in each T-Type. The number of galaxies in each T-Type is denoted on top of its corresponding violin. The region shaded in red depicts the early-type galaxies (ETGs: E, S0, and S0a), that are not part of this analysis, which concentrates on the types with the white background. The ETGs will be discussed in a different publication. The region marked in grey with the late-type spiral galaxies (Sdm and Sm) are highlighted due to their very low numbers. Additionally, these are visually harder to separate as two distinct morphologies, and will be merged together in certain plots in the analysis.} 
    \label{fig:btr_violin}
\end{figure*}
 
In order to study any trends related to morphology, we cross-match our catalogues with the MaNGA Visual Morphologies VAC (readers can refer to Sect. \ref{subsec:morph_cat}), which leads to sample of 1275 galaxies where we can study both mass and morphology trends.
Figure~\ref{fig:btr_violin} depicts the bulge-to-total ratio estimated from the $r$-band fluxes as a function of morphological T-Type and Hubble type (obtained from the VAC mentioned above) through violin plots. The violins are marked by a black box and whisker plot inside them, where the white circles denote the median $(B/T)_r$ for each type. The Sdm and Sm galaxies have very low numbers and wherever they are included in the analysis, it is important to note that those particular inferences are not statistically significant. Galaxies with irregular morphologies (Irr) and those with no classifications (NC) in the catalogue are excluded here. As would be expected, despite the large scatter in all morphologies, the median $(B/T)_r$ circles trace a trend starting with relatively high values for ellipticals and declining slowly with increasing type, flattening towards spirals. In the late-type spirals (Sd-Sm), our sample hits low-number statistics and begins to show an irregular trend. We note that all galaxies in our sample were run through \textsc{BUDDI} for a two-component decomposition prior to categorising their morphologies; therefore it appears that ellipticals have also successfully been decomposed into two components despite the traditional picture of a single (S\'ersic)-component model.  However, we expect it to be mostly an effect of forcing a two-component fit onto one-component objects. We note that from Sect. \ref{sec:galaxy_mah} onwards, our analysis is dedicated purely to spiral galaxies. Therefore, any possible physical motivation behind multiple components in elliptical galaxies will be investigated in future work.     

\subsubsection{Caveats on component stellar masses}
\label{caveats_masses}

There are some caveats with these derived component masses which should be discussed. Most importantly, these masses are derived, as explained above, by applying the structural parameters derived by the BUDDI fits on MaNGA data to the SDSS imaging data. However, the field of view of the MaNGA data is, by design, very small with an upper limit of $\sim2.5R_e$ (readers can refer to Sect.~\ref{subsec:manga_survey}), whereas it has been shown that a large field of view is critical in deriving accurate structural parameters, at least for components with high S\'ersic index values \citep{haeussler2007}. This means that the structural parameters from the BUDDI fits only really apply to the central parts of the galaxies, and not necessarily to their outskirts. This can have a large impact on the photometry of the components, for example it is known that the bulge profile (usually S\'ersic or de Vaucouleurs) dominates at large radii and obtaining an accurate sky measurement is vital to deriving good measurements of structural parameters \citep{haeussler2007}. Most importantly, the outskirts of a galaxy cannot be reliably taken into account in the absolute photometry, that is, the absolute mass scale can be somewhat compromised. Therefore, it should be remembered that the masses presented in this work for the bulges and discs represent the masses of the components that \textsc{BUDDI} identified and modelled as the bulges and discs, and thus may not correspond to the same components modelled in the same galaxies in other studies \citep{dominguez-sanchez2022pymorph, mendel2013quench}.

As a second effect, this of course also means that the photometry in general cannot represent the SEDs of the entire bulge or entire disc, potentially changing the bulge-to-disc mass ratio $(B/D)_{mass}$ or the bulge-to-total mass ratio $(B/T)_{mass}$ of the two components. Instead the component masses represent the bulge and disc as seen within the MaNGA field of view and assuming that their light profiles extend outside of this region in the same way. However, we have shown that our total masses agree well with masses derived by other means (Fig. \ref{fig:total-masses}). We emphasise that the main purposes of these derived masses is to examine global trends with component mass and not perform any detailed analysis with them. Therefore, the masses will not have an impact on the mass assembly histories themselves that are examined in this work (Sect. \ref{sec:galaxy_mah}), and such uncertainties on the component masses are deemed not to be critical here.

\subsection{Spectral fitting}
\label{subsec:spectral_fitting}

With the stellar masses of the bulges and discs in place, we now turn our attention to analysing the stellar populations of the different components in order to build a comprehensive picture of how spiral galaxies have assembled their masses throughout their lifetimes. A sub-sample of types Sa - Sm consisting of 968 spiral galaxies will form the basis for the rest of the analysis in this paper. We focus on extracting the mass-weighted stellar populations parameters through full spectrum fitting of the bulge and disc spectra of this BUDDI-MaNGA spiral galaxy sample using the penalised Pixel Fitting code \textsc{pPXF} \citep{CappellariEmsellem2004PASP, cappellari2017ppxf}. \textsc{pPXF} is an \textsc{IDL} and \textsc{Python}-based spectral fitting software that, when used in conjunction with a stellar spectral library, finds a linear combination of templates that best matches the observed spectrum after convolution with the line-of-sight velocity and velocity dispersion (LOSVD) of the galaxy. In this work, we opt for the \textsc{Python} implementation of \textsc{pPXF} v7.4.5 \footnote{Available from \url{http://purl.org/cappellari/software }}.

To fit the spectra, the MILES\footnote{Available from \url{http://miles.iac.es/pages/webtools.php}} evolutionary stellar population synthesis (SPS) \citep{vazdekis2015sps} models based on the BaSTI isochrones \citep{pietrinferni2004basti, pietrinferni2006basti, pietrinferni2013basti} were chosen as the stellar template library mainly owing to its rich metallicity coverage. The MILES library contains 985 stars with a wavelength coverage between 3\,500~$\AA$ and 7\,000~\AA, and a spectral resolution FWHM of 2.51~$\AA$ \citep{sanchez-blazquez2006miles}. We limit the fit to the spectral region between 4\,700~$\AA$ and 6\,700~$\AA$ (in the rest-frame), converting this range to the observed frame of each galaxy, such that it sufficiently covers the wavelength range between the $\mathrm{H\beta}$ and $\mathrm{H\alpha}$ lines. We normalise the input fluxes, and set a constant noise over the spectral range, defined by the reciprocal of the derived S/N. The grid formed by the fixed stellar population parameters range from -2.27 dex to 0.4 dex in 12 steps for the metallicities $\mathrm{[M/H]}$, and from 30 Myr to 14 Gyr in 53 steps for stellar ages, making up a total of 636 template spectra. We assume a Kroupa Universal IMF \citep{kroupa2001imf} with a slope of 1.3, and with no specific assumptions on the $\alpha$-abundance for the SPS models. In order to account for the shape of the continuum and correct for spectral calibration inaccuracies, a multiplicative Legendre polynomial of order 8 was adopted, excluding the need for explicitly specifying a reddening curve \citep{cappellari2017ppxf}. After masking out the [OI]5577 $\AA$ sky line, the stellar continuum is fit alongside the gas emission lines simultaneously. 

During the fit, we also employ regularisation to reduce intrinsic degeneracies to the fit solutions, that might otherwise lead to discrete weights and a bursty star formation history. The regularisation step allows for the smoothing of the variation of SSP weights with similar ages and metallicities. The level of smoothing is determined by the user-defined \textsc{regul} parameter in \textsc{pPXF}. While regularisation is useful in the case of individual spectral fitting and mostly recovers star formation histories that are physically interpretable (such as a continuous star formation episode as opposed to a discrete one with spurious features), it does not come without its limitations. \citet{shravan2015}  and \citet{norris2015} state that regularised fits do not prevent sharp jumps in the recovered star formation histories provided that such features are necessary to accurately fit the data, whether or not they are physically motivated. This situation may occur in cases where the spectrum is particularly noisy, leading to poor fits with the template spectra, or where the galaxy has undergone a small number of short bursts of star formation on timescales less than the difference in the age steps of the template spectra (readers can refer to \citet{zibetti2023} for a detailed study on constraining the minimum age resolution that can be achieved using SPS models). The challenge in regularisation lies in maintaining a balance between fitting the data and achieving smoothness in the solution, which can introduce potential biases \citep{cappellari2023} such as sudden shifts or jumps in the recovered SFHs. Therefore, some level of discreteness is not completely unavoidable especially in the above-mentioned cases, and we exercise caution while interpreting them (readers can refer to Sect.~\ref{subsubsec: MAH_individual}).

The degree of smoothing for each spectrum was determined following the approach outlined in \citet{cappellari2017ppxf} and described here. An unregularised fit with fixed kinematics is performed first, followed by a noise scaling step such that $\chi^2/N_{DOF}=1$ (DOF being the number of degrees of freedom in the fit). Then the fit is repeated over a range of user-defined regularisation values, and the optimal value is chosen when the increase in $\chi^2$, $\Delta\chi^2 = \sqrt{2\times N_{DOF}}$. For some galaxies, however, this does not constrain the optimal regularisation value well, and we opt the criterion following \citet{shravan2015}. The modified criterion is now $\Delta\chi^2 = (\sqrt{2\times N_{DOF}})/n$, where $n$ is an integer that we allow to vary between $1<n<100$. The constraint over $n$ that we have chosen is very liberal to allow fits to the spectra with very low S/N. We note that almost all of the component spectra with $n>20$ do in fact have low $S/N < 30$. We did not discard any fits because the fraction of objects with high $n$ values is very small compared to the entirety of the sample ($<2\%)$. It must be noted that the defined steps in stellar age in the synthetic template models are only an approximation as the models are based on averages of stars of similar ages and metallicities. In reality, star-formation is unpredictable and may occur on shorter timescales than defined in the models. Therefore, the resulting estimates derived from fitting these models might not be an absolute representation of the true star-formation activity in the galaxy components.

\subsubsection{Mass-weighted stellar populations.} 
\textsc{pPXF} estimates the mean metallicities and mean stellar ages in the stellar population analysis from the derived template spectra weights:

\begin{equation}
    \mathrm{log(age) = \frac{\Sigma\omega_{i} log(age_{temp,i})}{\Sigma\omega_{i}}}
\end{equation}

\begin{equation}
    \mathrm{[M/H] = \frac{\Sigma\omega_{i}[M/H]_{temp,i}}{\Sigma\omega_{i}}}
\end{equation}

 \noindent where $\omega_i$ is the weight of each ($i^{th}$) template, and $\mathrm{[M/H]_{temp,i}}$ and $\mathrm{age_{temp,i}}$ are the metallicity and stellar age of that same stellar template respectively. The weights $\omega_i$ in our analysis specifically represent the fractional contribution of each stellar template to the galaxy stellar mass $M_*$ (mass fraction). \textsc{pPXF} recovers the `formed stellar masses', which are associated with the zero-age mass distribution and do not include contributions from mass loss or supernova outflows (and are therefore expected to be higher than the present-day stellar masses).  

\subsubsection{Error analysis.}
\label{errors_analysis}

As in \citet{johnston2022buddi2}, we performed the errors analysis on the mean metallicities and stellar ages estimates by adding a random noise to the best fit spectrum obtained in the previous analysis, until it results in the same S/N as the original spectrum. This new simulated spectrum is fit with \textsc{pPXF} with the same noise scaling and regularisation value as before. This simulation was repeated 50 times for each spectrum, and the errors are taken as the 16th-84th percentile range of the resulting distributions of each stellar populations parameter.

\section{Galaxy mass assembly histories}
\label{sec:galaxy_mah}

With the mass weights obtained from \textsc{pPXF}, we can now reconstruct the mass assembly histories (MAHs) of the bulges and discs independently, by mapping out their cumulative mass assembly as a function of the lookback time. For this study, we turn our focus to the spiral galaxy sample of $1 \leq$ T-Type $\leq9$, which corresponds to Sa - Sm spiral galaxies. This sub-sample in BUDDI-MaNGA consists of 968 objects (Sect.~\ref{subsec:spectral_fitting}) with successful \textsc{pPXF} fits to their component spectra and stellar mass estimates. For the rest of the paper that focusses on spectroscopic properties, these galaxies form the basis for the analysis. In Sect.~\ref{subsec:downsizing}, we present our results on the individual mass assembly histories and the global trends we observe in the bulges and discs of the sample, as a function of their respective stellar masses. In Sect.~\ref{subsec:mah_mass_dependence}, we follow through with a better representation of mass-dependence using a mass-matched sample in every type. Finally in Sect.~\ref{subsec:mah_ttype_dependence}, we study the dependence of the morphological type on these mass assembly histories.

\subsection{Individual and global trends: Component downsizing}
\label{subsec:downsizing}

In the sections that follow, we first investigate the trends shown by the mass assembly histories of the bulges and discs of each galaxy in different morphological types. We then emphasise these trends through quantifiable parameters that indicate the formation mode of the components. Finally, we study the global trends observed on average by these mass assembly histories.

\subsubsection{Individual trends in bulge and disc mass assembly histories.}
\label{subsubsec: MAH_individual}
\begin{figure*}
    \begin{center}
        \includegraphics[ width=0.88\textwidth]{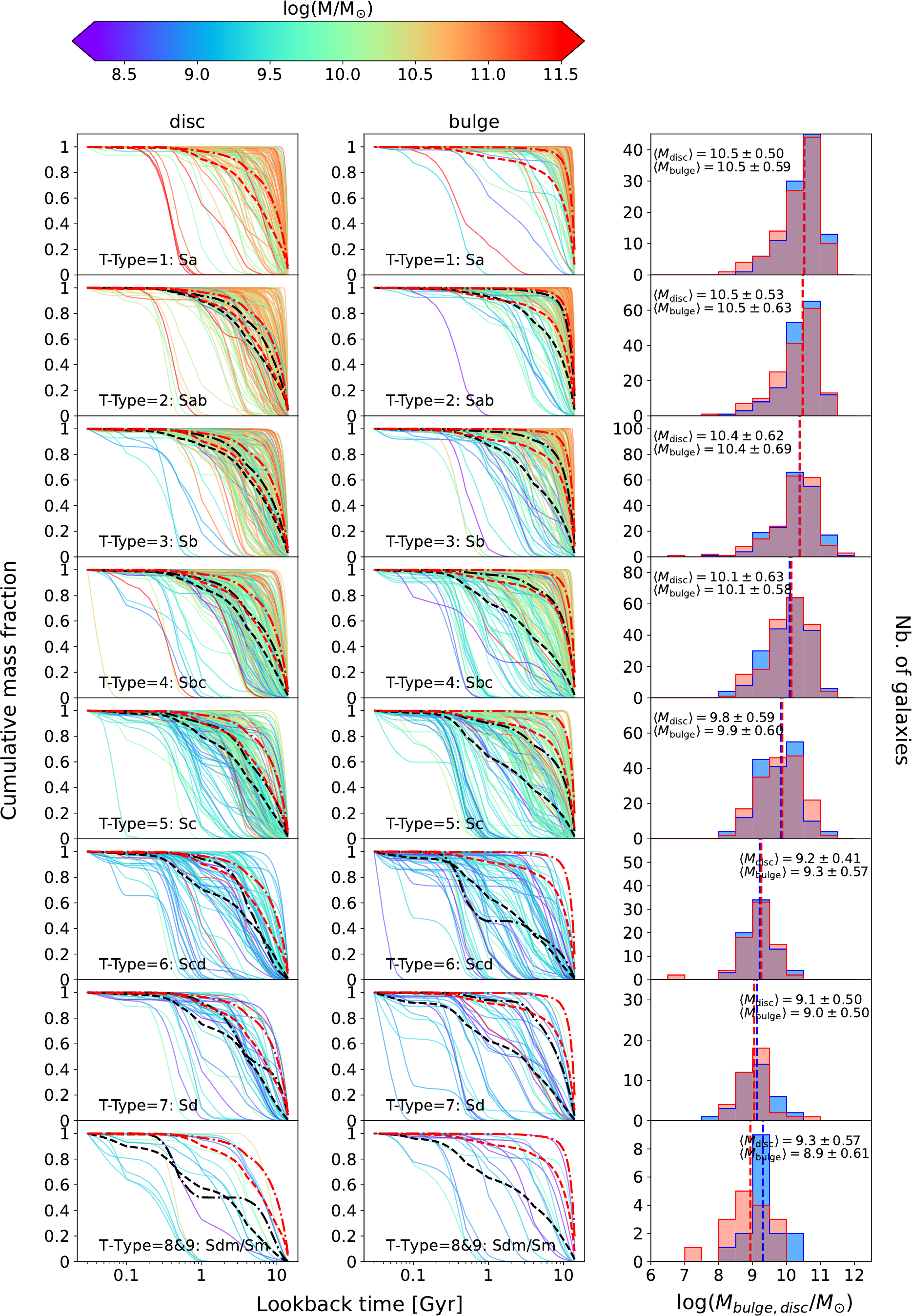} 
        
        \caption{\textbf{Individual and global trends of MAH in bulges and discs.} Left and centre columns: Individual MAHs of spirals separated by increasing T-Types (from top to bottom). In each panel, the mean-stacked MAH for high-mass components ($M>10^{10}M_{\odot}$) are shown as thick black dot-dashed curves, and low-mass components($M<10^{10}M_{\odot}$) as thick black dashed curves. These are depicted as red dot-dashed and dashed curves respectively for the Sa type (row 1), which also serves as a reference for all following types for easier visual comparison. Right-most column: Bulge and disc masses in red and blue, respectively, with their median masses shown by dotted lines. } 
        \label{fig:sfh_individual}
    \end{center}
\end{figure*}

The individual MAHs of the components are plotted in Figure~\ref{fig:sfh_individual}, in different bins of the T-Type: each row begins with the disc MAHs on the left, the bulge MAHs in the middle, and their respective stellar mass distributions on the right (blue for discs, red for bulges). The cumulative mass fraction assembled by each galaxy has been plotted against lookback time, which traces the star formation histories of galaxies across cosmic time. The cumulative mass fractions have been created by adding the normalised weights of the stellar spectra used in the \textsc{pPXF} fits at each age step from the oldest to the youngest stellar templates. Vertically, the plots step through increasing T-Types, starting with Sa in the top row, to Sdm and Sm in the bottom. The MAHs of the bulges and discs in the left two panels are colour-coded as a function of their respective stellar masses, with the highest masses in red and the lowest masses in blue and purple. The thick black dot-dashed and dashed lines represent the mean MAH in high ($M>10^{10} M_{\odot}$) and low ($M<10^{10} M_{\odot}$) component mass bins. These mean MAHs are shown in red only for the Sa type galaxies, since they act as a reference against which the MAH trends of the following types can be compared.

From the dot-dashed curves in Figure~\ref{fig:sfh_individual}, we observe that the more massive bulges assemble their stellar masses very early on within a short timescale, while the less massive bulges, for which the mean-stacked MAHs are shown as dashed curves, show a relatively delayed and longer-lasting mass assembly, an effect defined in \citet{cowie1996downsizing} as `downsizing' for galaxies as a whole. This effect is clearly visible even among the individual MAHs up to Sc type spirals, with a gradient from high masses in red and orange to low masses in green and blue - albeit the gradient being relatively stronger in bulges than in the discs. The morphological types later than Sc mostly host less massive bulges and discs (see the respective stellar mass distributions in column 3) with a large diversity in their MAHs, and a gradient is not clearly visible in either component for these. However, the large number of MAHs in these plots for each morphology makes the trend harder to see; for example, it can be made artificially weaker or stronger due to the number and order of the lines plotted. Especially in the latest T-Types, trends might become clearer when comparing average, or mass-matched samples. It must be noted that some of the individual MAHs appear to have sharp jumps over short timescales, which might imply extraordinarily high star formation rates seen in extreme starbursts. While this is a possibility, the discreteness can also be an effect of the limits of regularisation mentioned in Sect.~\ref{subsec:spectral_fitting}. However, since these sharp jumps occur only in a few objects relative to the total number of objects in our analysis, we do not expect this to alter our statistical results. Furthermore, in order to better identify any trends present, the next subsections will explore the data further by calculating and comparing the timescales over which half the mass was created in each component (Sect. \ref{subsubsec:half-mass}) and will consider the mean MAHs as a function of mass and morphology. 

\begin{figure*}
    \begin{center}
    \includegraphics[width=0.75\textwidth]{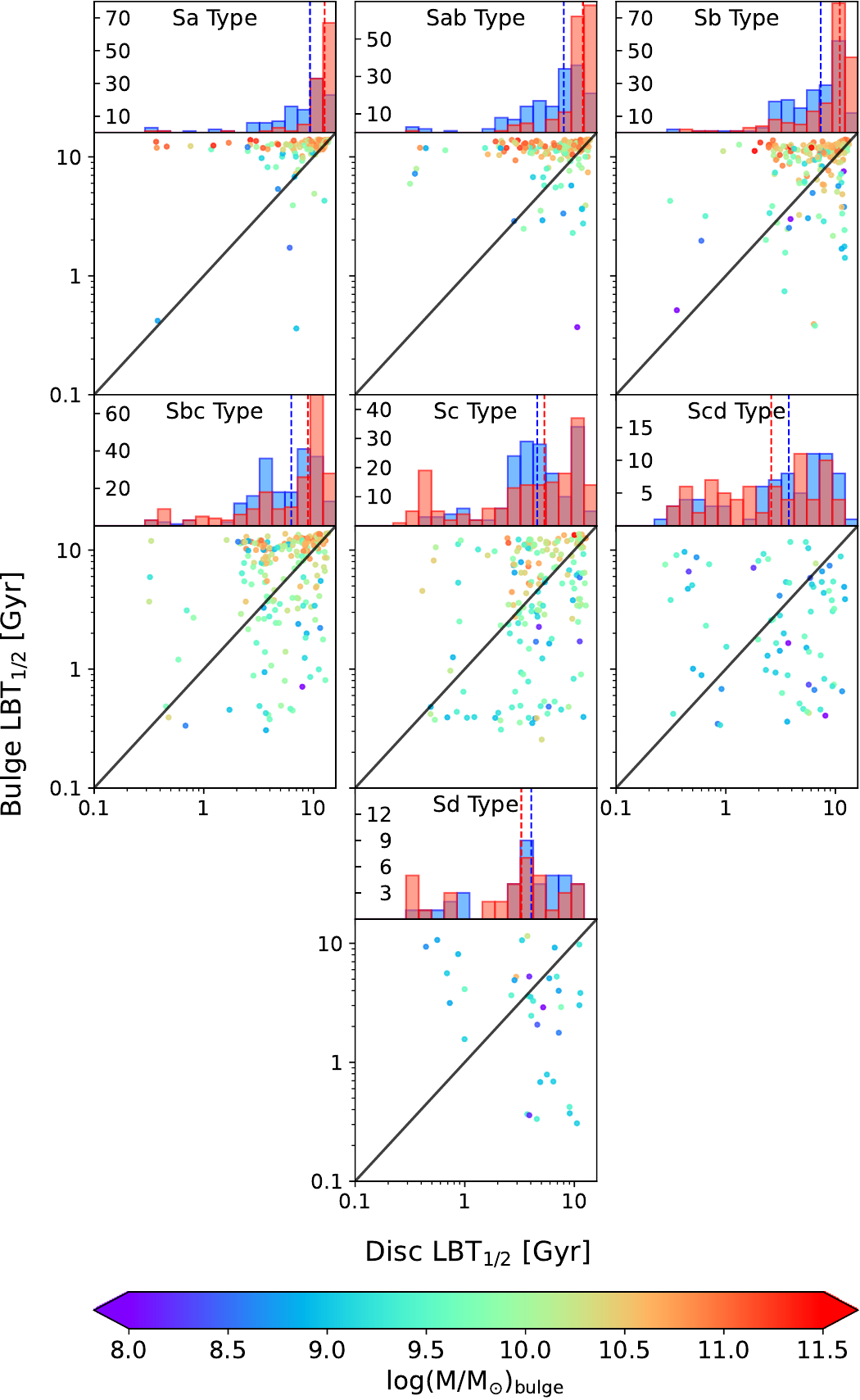} 
    \end{center}
    
    \caption{Formation time of the bulges and discs which marks the time it took to form 50 per cent of the stellar mass, colour-coded by the total stellar mass. The upper panels show the histogram of these formation times for the bulge and disc in red and blue respectively, with their median times in dashed lines.} 
    \label{fig:formation_time}

\end{figure*}

\subsubsection{Half-mass formation times and mass build-up.}
\label{subsubsec:half-mass}

We quantify the individual mass assembly histories through the \textbf{half-mass formation time} and a closely related parameter - the \textbf{half-mass formation timescale}. We define half-mass formation time as the lookback time ($\mathrm{LBT_{1/2}}$) when 50\% of the stellar mass had been built up. This parameter allows us to study which component assembled its stellar mass earlier. We define the half-mass formation timescale ($\tau_{1/2}$) as the difference between the lookback time when stellar mass assembly first occurred and when 50\% of the mass had been built up. This parameter provides information on which component assembled its stellar mass faster. Figure~\ref{fig:formation_time} shows the comparison of the formation times of the bulge and the disc for each galaxy, colour-coded by the total stellar mass of the galaxy. The solid black diagonal line marks the 1:1 correlation between them. The plots are split into seven panels in terms of their T-Types as in Figure~\ref{fig:sfh_individual}, excluding the Sdm and Sm types due to low-number statistics. The histograms in the upper panels show the $\mathrm{LBT_{1/2}}$ distributions of the bulge in red and the disc in blue, with their medians shown as the dotted lines. If bulge $\mathrm{LBT_{1/2}}$ $>$ disc $\mathrm{LBT_{1/2}}$, then the stars in the bulge assembled the majority of their masses before the disc, implying an inside-out assembly mode. Alternatively, galaxies with bulge $\mathrm{LBT_{1/2}}$ $<$ disc $\mathrm{LBT_{1/2}}$ would correspond to an outside-in formation mode, with the stars in the disc assembling the majority of its mass prior to the bulge. 

We can see a clear trend with Hubble type in this plot. Starting with the Sa spirals, the majority of the galaxies fall under the inside-out assembly regime (above the 1:1 line). Stepping horizontally across the different T-Types up until Sc, this result follows through, with the exception that the fraction of outside-in assembled galaxies increases with each type with the above-mentioned effect becoming less pronounced; nevertheless, the inside-out mode is still dominant. On reaching the Scd and Sd types, there does not appear to be any preferred assembly mode, and the galaxies are more or less equally distributed in both regimes. However, it must be noted that these types have relatively fewer galaxies in comparison to the earlier types, and it is possible this equal distribution is an artefact of low-number statistics. The histograms in the upper panels additionally show that the median bulge half-mass formation time is always higher than the median disc half-mass formation time for galaxies from Sa - Sc types. The Scd and Sd types appear to show a median outside-in assembly mode. However, the latter shows nearly identical half-mass formation time histograms for both components, and the time window between the bulge and disc $\mathrm{LBT_{1/2}}$ is quite narrow, suggesting instead that there was no preferred assembly mode.

With respect to the half-mass formation timescales, if bulge $\tau_{1/2}$ $<$ disc $\tau_{1/2}$, then the stars in the bulge assembled the majority of their masses in a shorter time span than the discs. On the other hand, if bulge $\tau_{1/2}$ $>$ disc $\tau_{1/2}$, then the disc stars assembled their masses faster than the bulge stars. From Figure \ref{fig:formation_timescale} in Appendix~\ref{appendix3}, we can see a correlation between the two parameters $\mathrm{LBT_{1/2}}$ and $\tau_{1/2}$. The same trends with morphology observed in Figure \ref{fig:formation_time} are mirrored in Figure \ref{fig:formation_timescale}, with the majority of Sa - Sc type spirals having assembled half of their bulge stellar masses faster than half of their disc masses. This fraction decreases, as we move from Sa towards Sc galaxies, whilst the fraction of galaxies showing the opposite trend increases. The later type Scd and Sd galaxies are equally distributed between those where the bulge stars assembled their mass faster than the disc stars, and those where the disc stellar mass assembly was faster. These trends are emphasised in the histograms (and the median bulge and disc $\tau_{1/2}$ lines) above each panel; additionally the time window between the bulges and discs is the highest for the Sa spirals, and becomes narrower with increasing type. These results tie in with our inferences from Figure \ref{fig:formation_time}, implying that a high fraction of bulges in Sa - Sc type spirals assemble half their stellar mass earlier (inside-out assembly) and in a shorter time span than the discs. The Scd and Sd types again suggest that the bulge and disc stellar masses were assembled together or have no preference in the assembly mode, with nearly equal assembly timescales. This is consistent with our inferences from the individual MAHs described in Sect. \ref{subsubsec: MAH_individual}.

\subsubsection{Global trends in bulge and disc mass assembly histories.}
To observe the mean global trends exhibited by the MAHs of the components (Fig. \ref{fig:sfh_individual}), the bulges and discs are split into two stellar mass bins at $10^{10}M_{\odot}$ (each by the component mass). The MAHs in each bin are then median-stacked, with the median of the galaxies with bulge and disc masses higher than $10^{10}M_{\odot}$ shown as the thick dot-dashed curve, and those lower than $10^{10}M_{\odot}$ shown as the thick dashed curve. The first row showing the Sa type spirals have the median MAHs marked in red, and point as a reference in each following panel with the different T-Types. For all other types, the median curves are shown in black, and can be directly compared to the earliest spiral type Sa, to more clearly see any trends with respect to Hubble type. Except for the latest Sdm and Sm type galaxies (bottom row), which do not contain high mass bulges, the downsizing trend is seen clearly especially in the bulges, where the MAH of both mass bins are very well separated, with the low mass bulges showing a slower mass build-up. On average, this downsizing appears to present itself in discs as well, albeit at a lower extent than the bulges. However, the high diversity observed in the MAHs requires a more narrow binning to better isolate the stellar mass dependence, that consequently also helps in matching the different T-Type sub-samples in their mass distributions (this will be addressed in the next section). 

On comparing the median (black) MAHs with respect to the Sa type (red), the bulges in both the high and low mass bins shift slowly to the left but progressively with type (up until Sd), indicating a relatively slower build-up of stellar mass. The Sdm and Sm types in the final row simply do not host any bulges over $10^{10} M_{\odot}$, but the shift in the low mass bin is still evident. The discs also show this shift, although the median separation between the low and high mass bins are much smaller than for the bulges, and are more delayed and extended as seen earlier from the individual MAHs. While these plots already provide us with a good deal of information, we must note from the stellar mass distributions in the different T-Types (shown in column 3 of Fig. \ref{fig:sfh_individual}), that they are not mass-matched. Additionally, there is a large scatter in the MAHs for which a simple binning into high and low masses might not provide an accurate representation of their assembly at all masses. In the following sections, this will be addressed to better understand the true mass-dependence of the assembly histories of bulges and discs.

\begin{figure*}
    \begin{center}
    \includegraphics[width=0.95\textwidth]{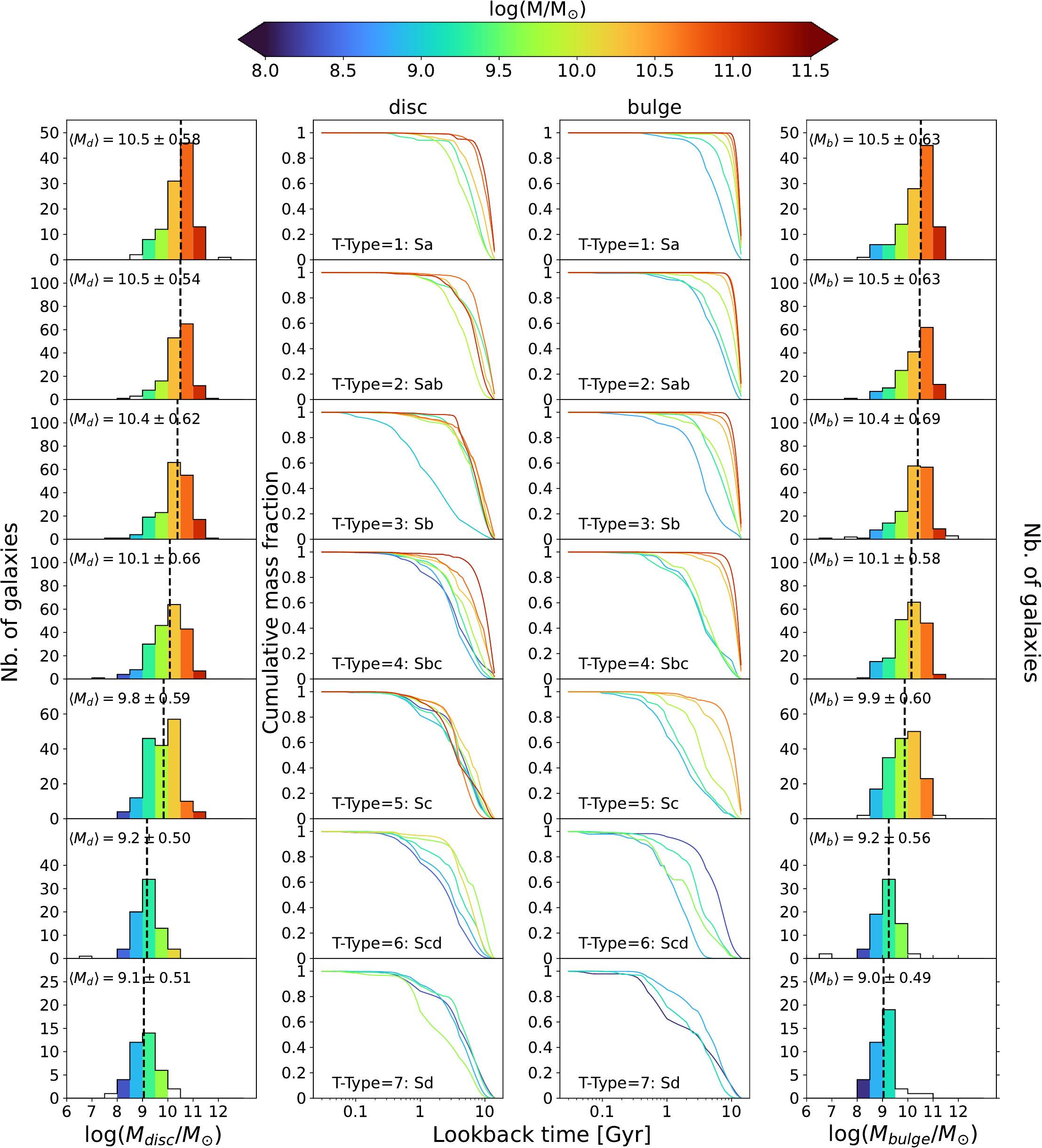} 
    \caption{\textbf{Dependence of MAH on component stellar mass.} Left-most column: Disc mass distributions in increasing steps of morphological type (top to bottom panels). Right-most column: Bulge mass distributions in increasing steps of morphological type. The vertical dashed line is the median component stellar mass for each type. Centre columns: disc and bulge MAH curves stacked in every corresponding mass bin shown in the histograms. Histogram bins and MAH curves have the same colour-coding, marked by the median mass of bulges or discs in each bin. Bins containing 3 objects or less are shown in white on the histograms, and their corresponding MAH curves have been excluded. }
    \label{fig:sfh_mass_dependence}
    \end{center}
\end{figure*}

\subsection{Dependence on component stellar mass}
\label{subsec:mah_mass_dependence}

While stellar mass trends can already be seen in Figure \ref{fig:sfh_individual}, the density of MAH curves in the plots make it difficult to really assess the strength of these trends. Additionally, 
given the disparity amongst the stellar mass distributions of bulges and discs for the different T-Types (see column 3 in Fig. \ref{fig:sfh_individual}), it is important that they are optimally matched to discard skewed or imbalanced inferences. In this section we will further investigate the effect of mass on the MAHs.

In order to look more closely at the dependence on the stellar mass of each component, we re-plotted Figure \ref{fig:sfh_individual}, but instead show the median-stacked MAH curves  in 0.5 dex bins of logarithmic stellar mass. These binned curves are shown in the middle two columns in Figure \ref{fig:sfh_mass_dependence}, with the colours representing the median component mass within that bin. The MAH of any bin with 3 galaxies or less was not plotted to avoid effects of small number statistics. The left and right-most columns in this figure show the mass distribution histograms for the discs and bulges, respectively, using the same mass steps and colour-coding for each mass bin. The galaxies excluded in the MAH plots (bins with 3 objects or less) are still shown in the component mass distribution histograms in white. The black dashed line indicates the median of the respective component mass for each type, showing that both the mean bulge and disc mass becomes lower,  moving from earlier to later type morphologies. This allows us to directly compare the stacked MAHs with their component distributions, and any type that does not have any objects in a particular bin simply does not show up in the plots in that corresponding colour. This allows our MAH plots to be mass-matched as we look at the curves of the same colour in different panels. 
 
Starting with the \textbf{Sa type} spirals in the first row, we see the aforementioned downsizing effect clearly in the bulges as a smooth gradient. The bulges of intermediate - high mass (orange-red) have assembled their stars very early on ($>8$ Gyr ago), and rapidly within a narrow timescale of the order of 1-2 Gyr. Moving to the low-mass bulges (green-blue), the mass assembly timescale increases successively, with them showing relatively prolonged star formation. Comparing to the discs in the same galaxies, this effect is still visible, but is strongly diluted. Overall, the most massive discs (red) do form first before the others; beyond this however, there is no consistent order to the disc assembly as a function of their mass. This result in both discs and bulges is seen again in \textbf{Sab type} spirals.

The downsizing trend is observed in \textbf{Sb type} spirals as well, but the bulges in the intermediate-high mass bins over $10^{10} M_{\odot}$ (orange-red) are more difficult to distinguish; they still do assemble earlier and faster than their less massive counterparts, however the order of assembly within these mass bins is less strict and harder to conclude. The discs might not show downsizing at the precision of the narrow mass bins, but the general trend of assembly still persists in the order of high-mass to intermediate-mass to low-mass discs (red-green-blue).

In the \textbf{Sbc type} spirals, the bulges continue to conform to this trend; additionally, at $\sim10^{10} M_{\odot}$ (light green), there is a clear separation between high and low mass MAH curves - the more massive bulges (red-orange) appear to have assembled their stellar mass rapidly through a single star formation event. The less massive ones (green-blue) have taken longer over the course of 10 Gyr with more than one star formation episode, showing more extended assembly histories. This result complies well with the global mass assembly histories from the previous section, but this is the last morphological type where the discs exhibit this trend.

The \textbf{Sc type} spiral bulges still show the downsizing trend for the intermediate-high mass bulges (orange-red), but for mass bins under $10^{9.5} M_{\odot}$, the MAH curves overlap and become harder to disentangle. The disc MAH curves are completely indistinguishable, where even the global downsizing effect begins to break down. Beyond this type, for the late \textbf{Scd} and \textbf{Sd} spirals, downsizing breaks down for both bulges and discs, and there appears to be no order in the stellar mass assembly of either component. These components show similar MAHs with longer delayed stellar mass assembly timescales with multiple star formation episodes. This might be a real result, or an effect of the faint bulges in these late-type spirals being technically difficult to recover. One must also note throughout these plots that objects below $10^9 M_{\odot}$ and above $10^{11} M_{\odot}$ are rare compared to the other mass bins (see left-most and right-most columns in Fig. \ref{fig:sfh_mass_dependence}), and although their MAH curves are still shown, they are not statistically significant. We find that these global results of the mass-assembly histories and their stellar mass dependence obtained through spectra are in accordance with a similar (albeit less detailed) analysis with SDSS photometry, using the star-formation history parameter $\tau$ (the e-folding time). These results are presented in Appendix~\ref{appendix1}.

The range of stellar mass assembly histories of the bulges and discs lends support to the idea that bulges residing in early and late-type spirals in the local Universe had been formed and grown through different mechanisms, and have a clear dependence on their stellar mass. The high-mass bulges in our sample always point to a fast and early mass assembly, which implies the formation of these entities through rapid or violent processes namely the monolithic collapse of a primordial gas cloud, or through major mergers. The low-mass bulges, on the other hand, are split between two cases: an assembly history that resembles the high-mass bulges (fast and early in Sa-Sab types), and one that resembles discs (slow or delayed, or both, with extended star formation episodes in Sb-Sd types). The latter suggests that the low-mass bulges were more likely formed from the discs through internal physical processes as well as environmental effects. While Fig. \ref{fig:sfh_mass_dependence} displays the median-stacked mass assembly histories that help us identify statistically significant trends, Fig. \ref{fig:sfh_individual} shows the signatures of both formation scenarios mentioned above in all morphologies, for each galaxy in the sample. It is also important to note that the presence of a composite-bulge system in some galaxies cannot be entirely discarded \citep{erwin2015bulges}, where the combined effects and properties are harder to disentangle, and could explain the range of mass assembly histories found in the bulges of spiral galaxies. Another theory that has been suggested in other studies is that a late-type galaxy bulge does not form purely through one of two scenarios but through a combination of both occurring on different timescales \citep{breda&papaderos2018}. A complement to this analysis by studying the stellar population properties of bulges and discs will be provided later in Sect. \ref{sec:stellar_pops}. In the line of the MAH analyses, the next step is to isolate the morphology trends, which is described in the following section. \\

\subsection{Dependence on morphology}
\label{subsec:mah_ttype_dependence}

\begin{figure*}
    \includegraphics[width=0.93\textwidth]{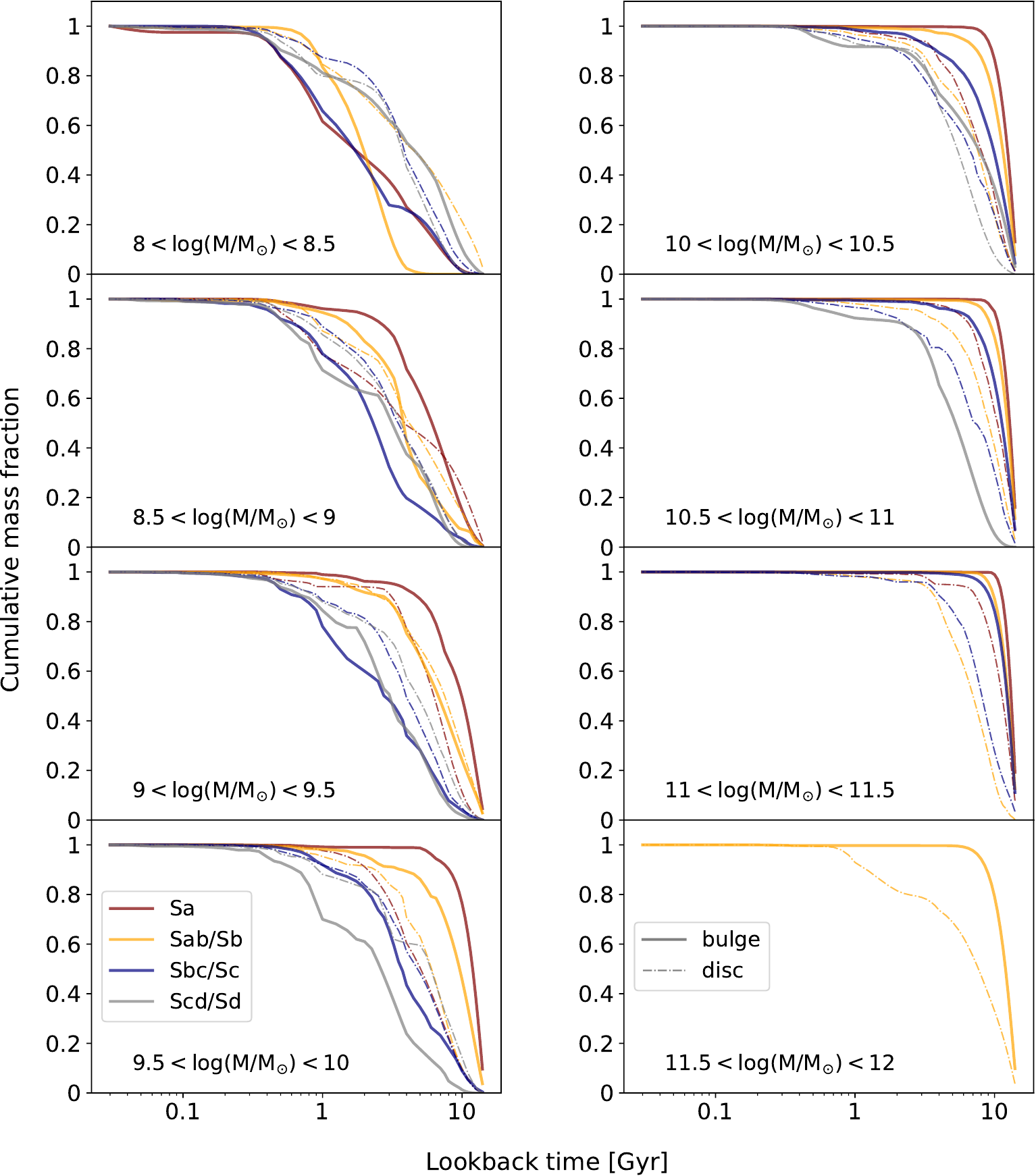} 
    \caption{\textbf{Dependence of MAH on morphology: } Each panel is a stellar mass bin defined in the previous section, increasing vertically from top to bottom. The morphologies were binned and stacked to show median MAHs of types Sa (maroon), Sab/Sb (orange), Sbc/Sc (navy), and Scd/Sd (grey). The solid thick lines represent the bulges and the thin dot-dashed ones correspond to the discs.}
    \label{fig:sfh_ttype_dependence}

\end{figure*}

With the component masses binned the same way as in Sect.~\ref{subsec:mah_mass_dependence}, we now separate the average mass assembly histories of the bulges and discs in each component stellar mass bin, as a function of morphological type (T-Type) in Figure~\ref{fig:sfh_ttype_dependence}. The bins increase in steps of 0.5 dex in logarithmic stellar mass vertically from top to bottom. In order to make the plots more readable and avoid over-plotting multiple curves, we bin the types into 4 major morphologies as Sa, Sab/Sb, Sbc/Sc, and Scd/Sd. The solid thick curves represent the bulge MAHs, and the thin dot-dashed lines represent the disc MAHs. In terms of morphological type, the Sa galaxies are marked in maroon, the Sab/Sb galaxies in orange, the Sbc/Sc galaxies in navy, and the Scd/Sd galaxies in grey. 

In the lowest mass bin ($\mathrm{8<log(M/M_{\odot})<8.5}$), the components do not appear to show any dependence with morphology - however, there are no Sa spiral discs in this mass bin. It might seem initially that despite a non-dependence on morphology, the discs have assembled their stellar masses prior to the bulges; however, this again might or might not be a skewed interpretation resulting from the fact that at these low masses, our sample simply has too few objects to statistically point to an outside-in assembly scenario.

The bulges do show a dependence on morphology - the Sa spirals consistently form first in all mass bins within $\mathrm{8.5<log(M/M_{\odot})<11}$, followed by Sab/Sb, later by Sbc/Sc, and finally by the Scd/Sd types. With increasing mass, we also note that the timescale between formation of bulges of different morphologies becomes shorter at each step. 

In the mass bin $\mathrm{9.5<log(M/M_{\odot})<10}$, the morphology dependence of discs starts to appear, although there is significant overlap for a majority of their assembly histories especially at earlier lookback times when the age uncertainties dominate. In the mass bin $\mathrm{10<log(M/M_{\odot})<10.5}$, this dependence becomes clearer and distinguishable between the early spirals and the late spirals (that is, the discs of types Sa-Sb have a clear earlier stellar mass assembly compared to types Sbc-Sd), although the morphology dependence is difficult to disentangle within these broader classifications. The dependence of the disc MAHs on morphology is best observed in the next two high-mass bins of $\mathrm{10.5<log(M/M_{\odot})<11}$ and $\mathrm{11<log(M/M_{\odot})<11.5}$, where the curves can be easily discerned and follow the same assembly order as the bulges as described earlier (Sa first, followed by Sab-Sb, then by Sbc-Sc). We note at these high masses of $\mathrm{log(M/M_{\odot})>10.5}$, our sample lacks components of the late Scd-Sd types, but nevertheless emphasise that still shows the same order of bulge and disc mass assembly in terms of morphology, for the ones that do exist. The final mass bin $\mathrm{11.5<log(M/M_{\odot})<12}$ only contain galaxies of types Sab-Sb which show the general trend of an earlier bulge stellar mass assembly compared to the disc by several Gyr.    

The downsizing trend is visibly justified looking at the decreasing assembly timescale with increasing mass steps, as the lines get steeper with increasing mass. The morphology dependence is evidently stronger for the bulges than the discs, where it only becomes apparent at higher masses.

These results on the dependence of the mass assembly histories of bulges and discs on their respective stellar masses and morphologies, show that the primary driver is indeed the stellar mass. While the morphology definitively plays a role in determining the assembly mode, its dependence is relatively less substantial.

\subsection{Effect of masking foreground stars}
\label{subsec:star_mask_effect}

\begin{figure*}
    \includegraphics[width=\textwidth]{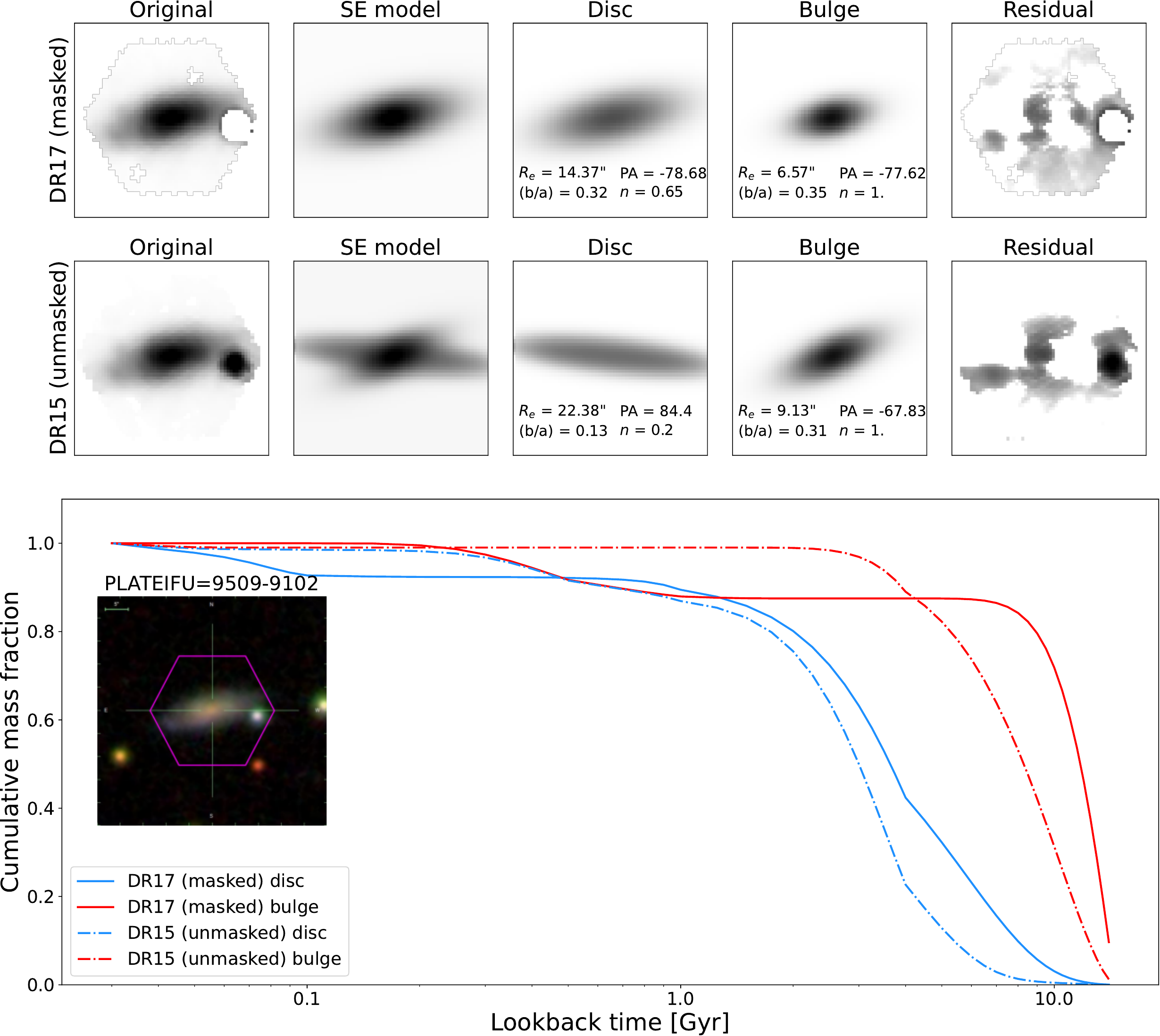} 
    \caption{Bulge-disc decomposition of galaxy 9509-9102 from the previous BUDDI run on DR15 (second row) and current DR17, which includes a foreground star mask (upper row). The final panel shows the MAH of the bulge (red) and disc (blue) resulting from both fits (solid curves for DR17 and dot-dashed curves for DR15), along with an SDSS $gri$ image of the galaxy with the MaNGA field of view (91 fibre IFU) superimposed in purple.}
    \label{fig:star_mask}

\end{figure*}    

A significant improvement in the current DR17 version of \textsc{BUDDI} fits over the previous DR15 ones is that we have incorporated the masking of foreground stars in the field of view of the target MaNGA galaxies. In the fitting process with \textsc{GalfitM} (and by extension \textsc{BUDDI}), the magnitudes and structural parameters of the target galaxy can be altered due to the presence of a foreground star or a neighbouring galaxy if those are not taken into account properly. The Data Reduction
Pipeline \citep[DRP;][]{law2015drp} of all SDSS-MaNGA data releases includes foreground star masks that were created through visual inspection by MaNGA team members. From DR15 however, these star masks were augmented by those created by citizen scientists in the Galaxy Zoo: 3D project \citep{masters2021gz3d}. For more details on the project and the criteria used in identifying and creating foreground star masks, we refer the reader to Sect. 3.2 in \citet{masters2021gz3d}. The masks were created as a circle with a 2.5\arcsec radius (MaNGA PSF) around the clustered location marked by the volunteers. These masks identify and mask all stars that are visually deemed important enough to influence the output of the \textsc{GalfitM} fit significantly.     

Figure~\ref{fig:star_mask}  compares the masked and unmasked fit results and mass-assembly histories for the galaxy with MaNGA plate-IFU 9509-9102, which has an Sab type morphology. The upper row depicts the final \textsc{BUDDI} fits after the foreground star within the hexagonal field of view has been masked out, and the middle row shows the fits from the previous run where the star was included in the fitting process. The original datacube image is shown in the first column, followed by the combined S\'ersic + Exponential model, the disc model, the bulge model, and finally the residual. All the images are collapsed along the wavelength direction to create the median white-light image. The fit parameters indicating the effective radius ($R_e$), the position angle (PA), the axis ratio ($b/a$), and the S\'ersic index ($n$) for each component is shown in their respective panels. It is immediately apparent from these values, as well as the bulge model and the fit residuals, that the masking of this bright nearby star is essential to derive physically meaningful fit parameters, and consequently the spectra and MAHs. To show the effect of this mask on the MAHs, we compare the MAHs of the bulges and discs before and after masking out the star in the fits, in the lower-most panel in Figure \ref{fig:star_mask}. The solid blue and red curves indicates the MAHs of the disc and the bulge respectively, where masking is included, and the dot-dashed blue and red curves represent those where the star was not masked out. On the left, we also show the SDSS $gri$ image of the galaxy with the hexagonal field of view superimposed on it.

From the disc model of the DR15 fit, it is clear that in the unmasked fit, \textsc{GalfitM} has tried to include the neighbouring star in the fit, resulting in an unusually extended light profile for the component, with a S\'ersic index of 0.2 (which is also the lower limit of the fits) and $R_e$ of 22.38\arcsec (which is the most unfeasible and significantly affected parameter). Moreover, the disc axis ratio ($b/a=0.13$) suggests a highly edge-on galaxy, which is not the case as seen from the SDSS image and the MaNGA datacube. Since the foreground star appears very close to the disc, the fit has assumed it to be a significant part of the disc and has completely shifted the position angle of the disc in order to incorporate it. The bulge has also been modelled as an elongated component (with a large $R_e$ of 9.13\arcsec) that has incorporated the light from the true disc component of the galaxy.   

In the new DR17 fit with the star masked out, \textsc{GalfitM} has modelled the bulge and the disc cleanly with a more physically motivated S\'ersic indices of 1 and 0.65 respectively. The fit parameters of the discs are significantly changed compared to the DR15 fits. The position angle of the bulge and the disc have similar values as would be expected, as opposed to the DR15 fit, where they were oriented in opposite directions. The $R_e$ for the bulge and disc are also physically feasible with values of 6.75\arcsec and 14.37\arcsec respectively. Additionally, the estimated axis ratio for the disc is higher ($b/a = 0.32$) than for the unmasked fit, indicating the galaxy is not as edge-on as previously modelled. Finally, the combined SE fit has managed to cleanly model the light profile of the entire galaxy this time. 

On comparing the MAHs between the two versions of the fits, we can see that although the general mass assembly order has not changed, they trace substantially different star formation histories. Moreover, through the unmasked fits, the stellar populations of both the bulge and disc are biased towards relatively younger stellar populations up to $\sim1$ Gyr ago. With both the masked and unmasked fits, it appears that the bulge started assembling its mass first within a few Gyr after the Big Bang. The disc on the other hand, has had a more delayed and slower star formation starting after the bulge. However, for the unmasked fits, the bulge mass assembly history is more prolonged compared to one from the masked fits. Additionally, while the unmasked fits point to a single episode of star formation that built up the entire stellar mass of the bulge within a few Gyr, the masked fits instead show more recent star formation, as late as 0.3 Gyr ago. The disc mass assembly histories are similar for both the masked and unmasked fits up to a lookback time of $\sim0.5$ Gyr, after which the new MAH suggests continued star formation leading up to current times, while the old MAH indicates that the stellar mass has already been built-up soon after. 

The foreground star which is present at a different redshift than the target galaxy, adds a strong superimposed spectrum onto the disc spectrum due to their apparent proximity. The spectral features of the star would therefore appear in the disc spectrum at the wrong positions and can lead to a poor fit with \textsc{pPXF}. However, it must be noted that this example is not a global representation of all the galaxies in the sample with a foreground star, and that the results change depending on the apparent proximity of the star to the galaxy. We have visually inspected the effects of masking the star for $\sim 100$ randomly chosen objects from the GZ:3D catalogue, and find that in several cases where the star is at a sufficient distance from the bulge and the disc, even if it is present within the MaNGA field of view, only has a negligible effect on the MAHs. However, we rarely find cases where masking the star leads to a worse fit or poses a disadvantage. As such, we have decided to use the masked fits in all galaxies where they exist, for consistency. We observe that when the foreground star is in fact in close proximity to the galaxy such that it can affect the spectrum of either component, masking it out clearly improves the results of the fits. This agrees with the findings in \cite{haeussler2007}. Therefore, masking foreground stars while fitting the galaxy components ultimately provides more accurate and physically motivated stellar populations and star formation histories.

\section{Stellar population properties of bulges and discs}
\label{sec:stellar_pops}

\begin{figure}
    \centering
    \includegraphics[width=0.9\columnwidth]{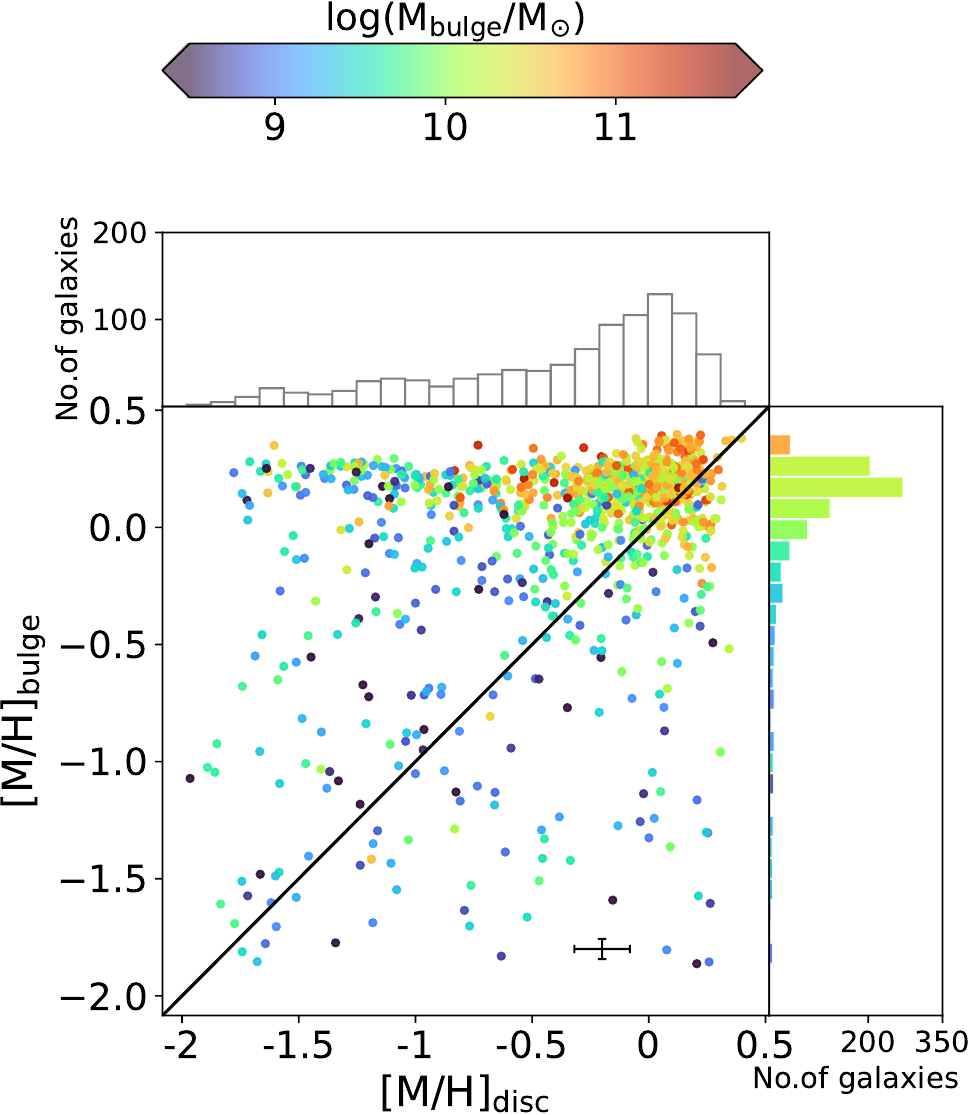}
    
    \includegraphics[width=0.9\columnwidth]{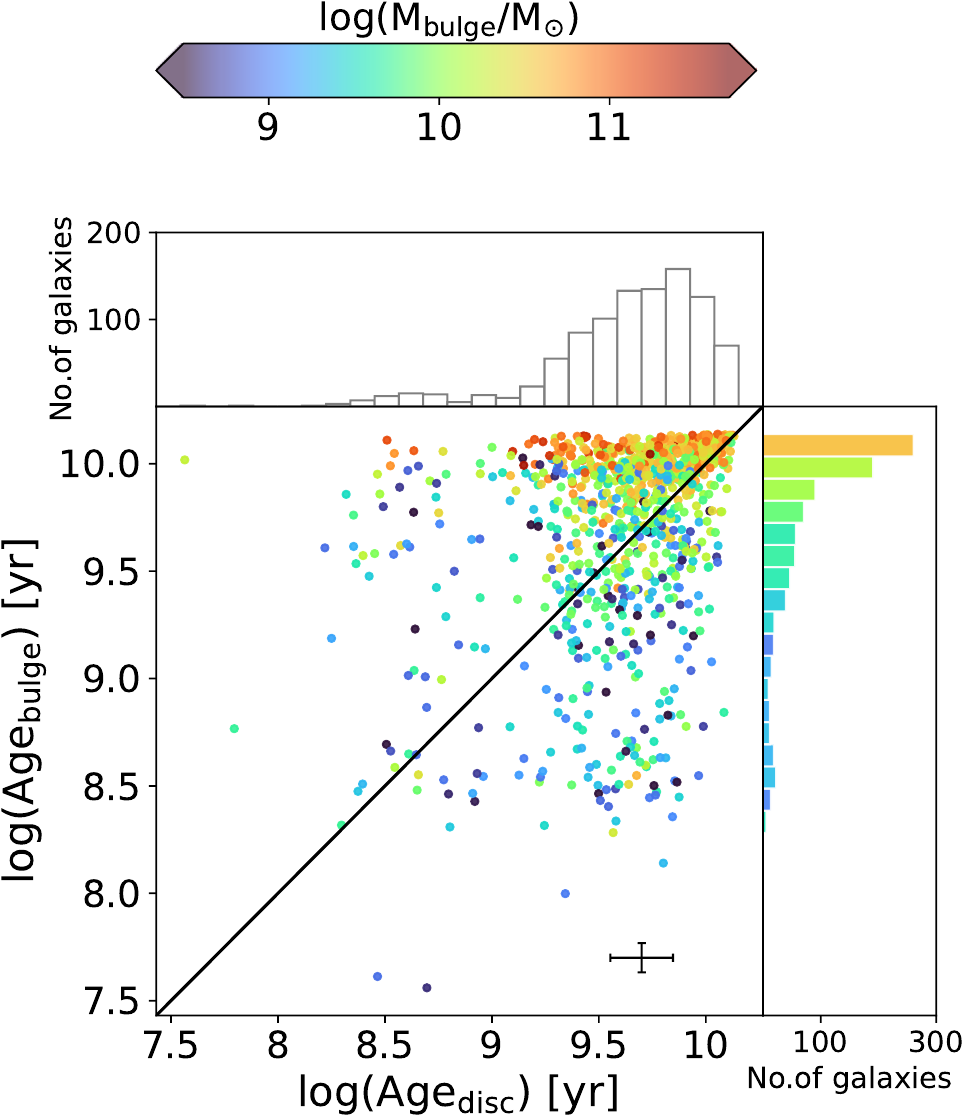}    
    \caption{Stellar population properties of bulges and discs along with their corresponding distributions. Upper panel: Comparison of mean mass-weighted stellar metallicities of the bulges and the discs, colour-coded by bulge stellar masses. The distributions in bulge (disc) metallicities are shown in the right (upper) joint histograms. Each bin in the histogram of the bulge metallicities is colour-coded by the mean bulge mass in that bin. Lower panel: The same as above, but for stellar ages. }
    \label{fig:met_age_trends}
\end{figure}

As described in Sect.~\ref{subsec:spectral_fitting}, the mean mass-weighted stellar population properties of the bulges and discs were derived from \textsc{pPXF} after fitting the `clean' spectra. In this section, we explore the dependence of these stellar population properties on stellar mass and morphology. In Figure~\ref{fig:met_age_trends}, the upper panel compares the mean stellar metallicities of the bulges and the discs, colour-coded by the bulge mass. The error bars shown denote the median uncertainties for the sample as computed in Sect. \ref{errors_analysis}. The black diagonal line marks the 1:1 correlation between them - above this line, the bulges have higher metallicities than the discs. The joint histograms show the distributions of the component metallicities (disc [M/H] in the upper-joint panel and bulge [M/H] in the right-joint panel). Furthermore, each bin in the bulge metallicities histogram is colour-coded by the median bulge mass in that bin, which allows us to highlight trends that might be hidden beneath the volume of points. The lower panel compares the mean logarithmic stellar ages of the bulges and discs in a similar way. 

\begin{figure*}[h!]
    \centering
    \includegraphics[width=0.9\textwidth]{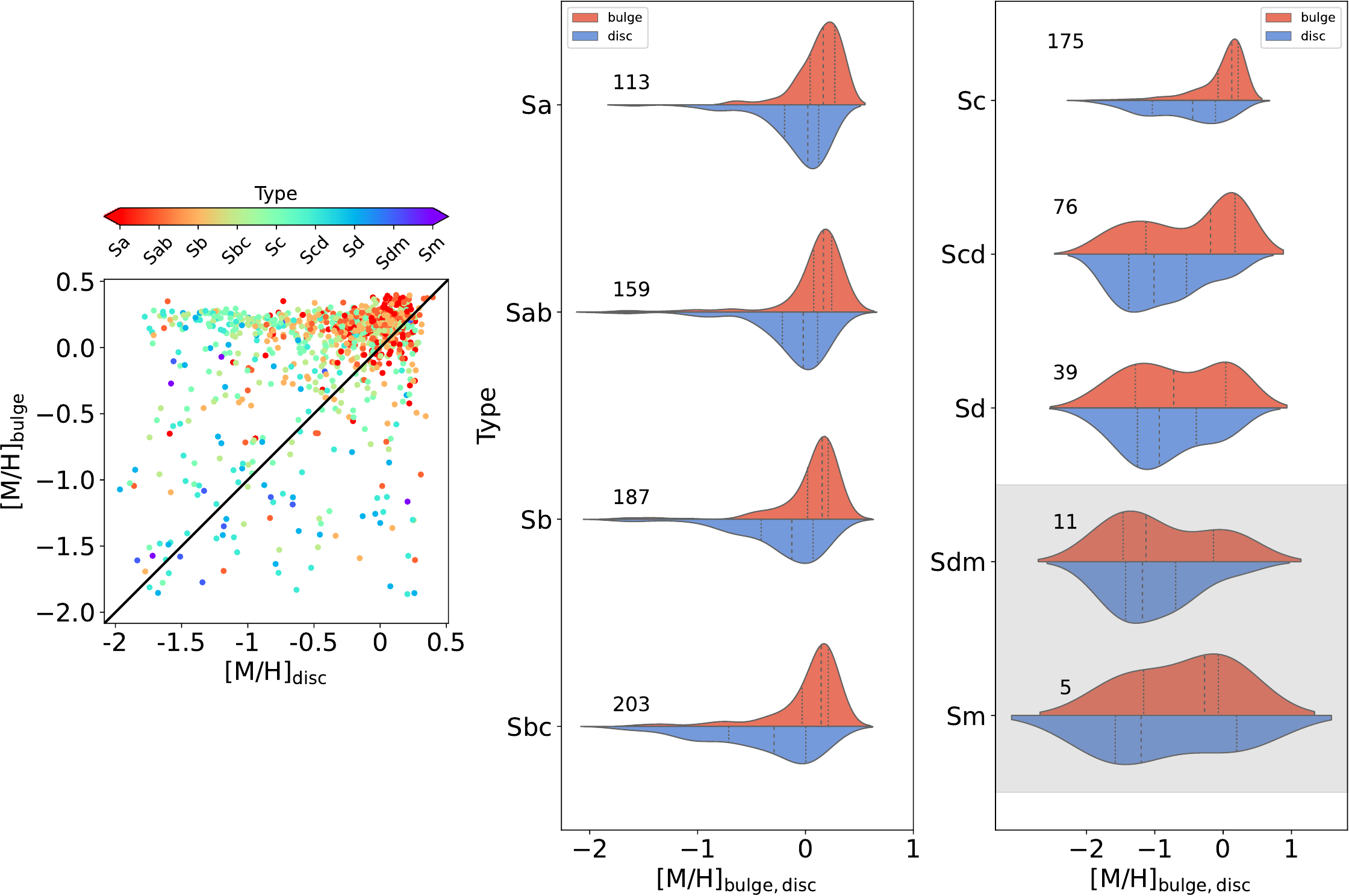}
    \caption{Comparison of bulge and disc metallicities and their dependence on morphology. Left panel: Bulge [M/H] compared to disc [M/H], colour-coded by the morphological type. The next two columns show split violin plots, with the upper half depicting the bulge metallicity distribution (red) and the lower half depicting the disc metallicity distribution (blue) in each type. The violin plots are marked with the median by the dashed black lines, and the quartiles are shown by the dotted lines. The grey shaded regions show the morphologies with very low-number statistics. }
    \label{fig:bulge-disc-met-violin}
\end{figure*}

\begin{figure*}[h!]
    \centering
    \includegraphics[width=0.9\textwidth]{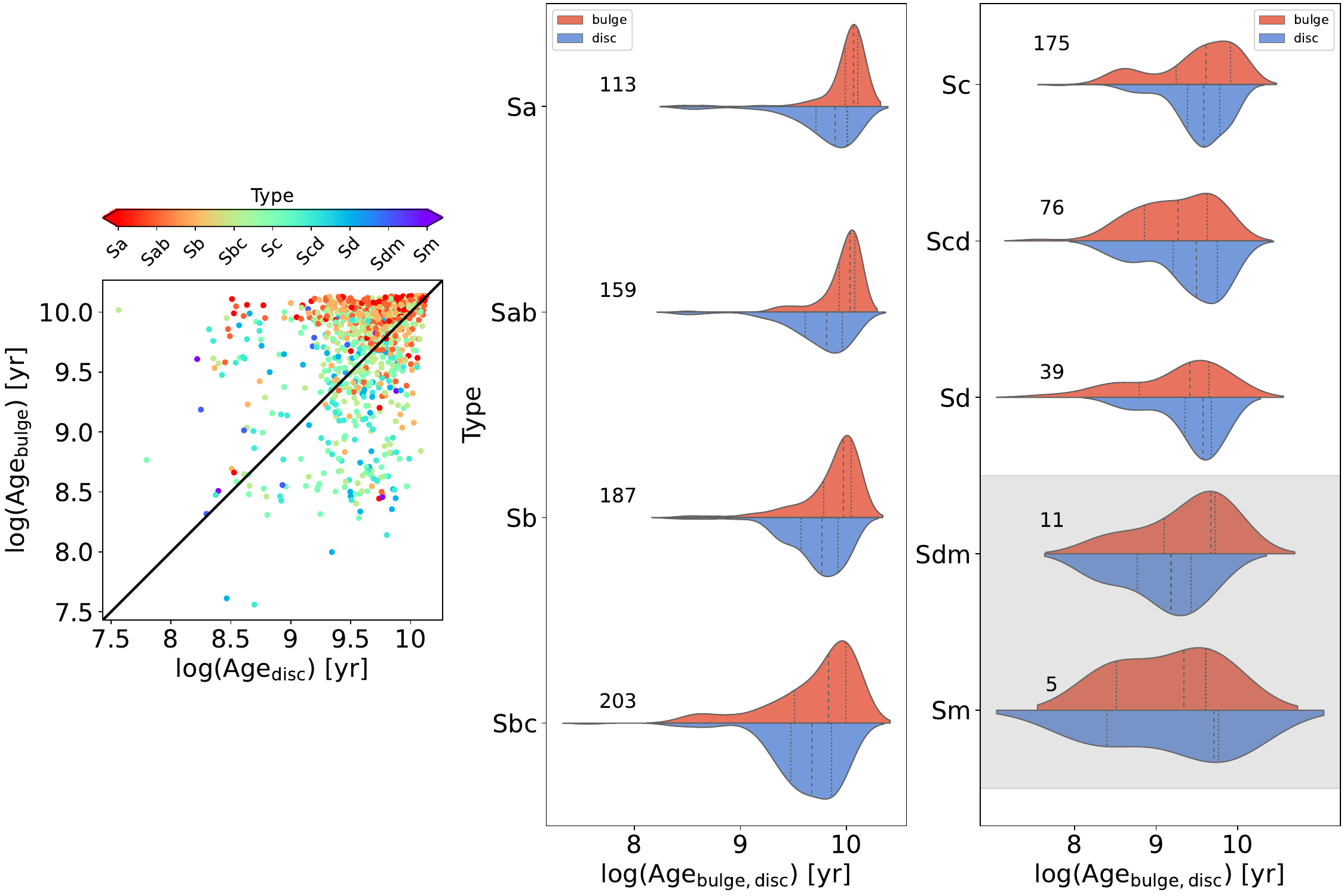}
    \caption{Comparison of bulge and disc stellar ages and their dependence on morphology. Left panel: Bulge $\mathrm{log(Age)}$ compared to disc $\mathrm{log(Age)}$, colour-coded by the morphological type. As in Figure \ref{fig:bulge-disc-met-violin} next two columns show split violin plots, with the upper half depicting the bulge age distribution (red) and the lower half depicting the disc age distribution (blue) in each type. The violin plots are marked with the median by the dashed black lines, and the quartiles are shown by the dotted lines. The grey shaded regions show the morphologies with very low-number statistics.  }
    \label{fig:bulge-disc-age-violin}
\end{figure*}

Figures \ref{fig:bulge-disc-met-violin}
and \ref{fig:bulge-disc-age-violin} compare the stellar population properties of bulges and discs ([M/H] in left panel of Fig. \ref{fig:bulge-disc-met-violin}; ages in left panel of Fig. \ref{fig:bulge-disc-age-violin}), colour-coded with respect to morphological type (early types in red to late types in blue). The violin plots in the next two panels study the stellar populations of the bulges and discs of each morphological type. The upper violins in red show the [M/H] distributions (Fig. \ref{fig:bulge-disc-met-violin}) and stellar age distributions (Fig. \ref{fig:bulge-disc-age-violin}) of the bulges, while the lower violins in blue show those of the discs.

\subsection{Trends in mass-weighted stellar metallicity}

From Figure \ref{fig:met_age_trends} (upper panel), for the whole spiral galaxy sample, we find that the bulges are in general more metal-rich. On average, we find the metallicity difference between the bulge and disc within the same galaxy to be $\sim0.34$ dex. The high-mass bulges of $\sim \mathrm{M_{bulge}>10^{10} M_{\odot}}$ (red-orange points) almost all cluster above the 1:1 line, indicating these form the highest fraction of galaxies with bulges more metal-rich than discs. With decreasing bulge mass, the metallicities also appear to become lower, with those for the lowest masses of $\mathrm{M_{bulge}<10^{9} M_{\odot}}$ (blue points) showing a broad range in metallicities. Below the 1:1 line, the majority of points show lower bulge masses. This trend is further highlighted by the right-joint histogram, where we can see that the bulge stellar metallicities correlate well with bulge stellar masses. The general trend is clear starting with the high-mass bulges (orange) on the high-metallicity end, towards the low-mass bulges (blue) on the low-metallicity end. These results emphasise the dependence of bulge stellar metallicities on the bulge mass, with the more massive bulges hosting more metal-rich stellar populations than the low-mass counterparts. 

In Figure \ref{fig:bulge-disc-met-violin}, the left panel shows that the majority of spirals (Sa - Sc types from red to light-blue) have the most metal-rich bulges and discs, and the bulges are significantly more metal-rich than the discs. However, due to the density of points and their plotting order, much of the trends are hidden. The violin plots in the figure provide a clearer depiction of the morphology dependence by splitting the metallicity distributions of the bulges and discs for each type. We note that the results mentioned above are reproduced in the violin plots. For spiral galaxies from types Sa - Sc, the bulges are more-metal rich than the discs (shown by the median lines), with their difference increasing towards the later spiral galaxies (towards Sc). The discs start showing a wide range of metallicities and a clear tail towards lower values. For types later than this, the bulge metallicity distributions show a bimodality with peaks in the higher and lower [M/H] ends, although for the very late types this result is limited by very low-number statistics (Sdm-Sm). However, the median trend still follows - although the bimodality in the bulges of late-type spirals is significant. 

It can be seen in Figure \ref{fig:met_age_trends} that $\sim16\%$ of the discs and $\sim7\%$ of the bulges of the spiral galaxies appear to have [M/H] $< -1$ dex. Such low metallicity ranges are more comparable to those found in low-mass galaxies ($M_*<10^9M_{\odot}$) than in spirals. There could be several potential reasons for these low metallicity estimates from \textsc{pPXF}. For example, in some cases, the S/N of the decomposed spectra are less than 20, which makes stellar population analysis with full spectral fitting more difficult \citep{zibetti2023}. One potential cause of these noisy spectra could be particularly low or high bulge-to-total ratios. In this work, we considered the fits to be successful if the magnitude difference of the two components was between $0.1\leq B/T_r\leq0.9$. In photometric bulge-disc decomposition, the B/T light ratio is often selected to be more conservative \citep{haeussler2013megamorph, nedkova2021mzr, haeussler2022galfitm}, where they recover very reliable structural parameters within $0.2<B/T<0.8$ . While we justify our lower limit by using information from the entire wavelength range to derive a reliable fit, there can still be extreme cases where either the bulge or the disc component might be particularly faint, thus making it harder to measure the stellar populations reliably. Another effect that should be mentioned is the age-metallicity-extinction degeneracy. While full spectral fitting with \textsc{pPXF} is often considered to reduce this degeneracy relative to photometric colours or line strength indices, it may not remove the effect completely. Since the sample of galaxies presented here have spiral morphologies, a significant fraction of them likely contain dust discs. These dust discs might not be easily resolvable in the MaNGA data, but they could still affect the decomposed spectra by leading to artificially low values of metallicity. The above-mentioned effects can dominate individually in either component in the spectral fitting process or combined to a point that they cannot be disentangled. Therefore, we add a word of caution towards interpreting these estimates as absolutes. Even so, the general statistical trends we have highlighted hold despite the caveat.

\subsection{Trends in mass-weighted stellar age}

From Figure~\ref{fig:met_age_trends} (lower panel), the bulges are in general older than the discs. On average, we find the stellar age difference between the bulge and disc within the same galaxy to be $\sim 1.7$ Gyr. We see a clear trend with respect to the bulge ages, with the most massive bulges (red-orange points) being the oldest, which again cluster together above the 1:1 line. Moving towards intermediate and low-mass bulges, their ages become younger as well. This trend is emphasised in the right-joint histogram, starting with the high-mass bulges (orange) that are older, towards the low-mass bulges (blue) that are younger. The dependence of the stellar populations on bulge mass is stronger for the stellar ages than for the metallicities as shown in the histogram. 

In Figure \ref{fig:bulge-disc-age-violin}, the left panel shows clearly that the early spirals from Sa - Sb have the oldest components, and host bulges that are older than their corresponding discs. However for the later types from Sbc - Scd, the points appear to be spread out across a range of ages. There is a significant number of galaxies that lie below the 1:1 line, indicating the presence of a second population with discs older than their bulges. This result again follows through in the different T-Types shown by the violin plots in the next two panels. For types Sa-Sc, the bulges are consistently older than the discs on average, as shown by the median lines. Beyond this, the trend seems to shift with discs appearing to be older, until the very late-types where we hit too low numbers to reach conclusive results. The bulge distributions fall steeply off at the higher end of stellar ages, while showing a tail towards lower values. The discs show a broader range in stellar ages than the bulges (except for Scd and Sd types), reflecting their different, more extended star formation histories compared to the bulges.  
However, we note that the stellar age determination beyond 8 Gyr has been often found to be highly uncertain \citep{ibarra-medel2016, ibarra-medel2019}.

Although we do not show the dependence on disc mass in the plots, we find very similar general trends with more massive discs hosting older and more metal-rich populations compared to their low-mass counterparts. Finally, these trends in stellar population properties of the bulges and discs with both stellar mass and morphology reflect different formation mechanisms for each component. For instance, a high fraction of bulges are old and have high masses, and appear to have assembled the majority of their stellar masses early on in their lifetimes. However, a smaller but significant fraction of bulges and discs show the opposite trend, or show similar stellar populations in both, reflecting a slower mass assembly in these components. The possible reasons for this will be explored in the discussion.

\section{Discussion}
\label{sec:discussion}

In the series of papers based on BUDDI-MaNGA, Paper I \citep{johnston2022buddi1} introduces the project and describes the technical details behind modelling bulges and discs of galaxies from IFU datacubes in order to cleanly extract their spectra. Consequently, it provides an overview and characterisation of the first fits to the BUDDI-MaNGA sample built from DR15, along with statistical and reliability tests. Paper II \citep{johnston2022buddi2} presents the first scientific results centred on the S0 galaxies in the BUDDI-MaNGA DR15 sample. It probes the stellar populations of the bulges and discs in these galaxies, and how their star formation histories trace their origins and evolutionary pathways. In this paper, we further build on our sample size with the final release of SDSS-MaNGA DR17 and improve the fits by masking the neighbouring foreground stars. We have nearly doubled the number of galaxies in this latest version to 1452 objects (with SE fits for all morphologies available) in the final sample, and undertake a follow-up study by tackling the sample with the highest morphological frequency: the spiral galaxies. Through full spectral fitting with \textsc{pPXF}, the mass-weighted stellar metallicities and stellar ages of the two individual components are estimated for 968 spiral galaxies, along with their mass assembly histories, to help unravel a picture of their formation and evolution. It must be noted that in both versions of the fits, there is an inherent selection bias towards the high-mass galaxies because our initial sample selection criterion is to use only those galaxies that were observed with the two largest IFU sizes (91 and 127 fibre IFUs), which consequently does not sample the low-mass systems \citep{johnston2022buddi2} as well as the high-mass ones (see Fig. \ref{fig:overview}).     

Our analyses and results target the mass-assembly histories of bulges and discs as a function of their respective stellar masses and morphology, emphasising the trends they exhibit individually as well as globally. We complement these results by analysing the properties of stellar populations hosted by bulges and discs and similarly their dependence on stellar mass and morphology. Throughout this paper, we only focus on the mass-weighted stellar populations, which trace the evolutionary history of the galaxies and are not badly affected or biased by recent star formation. In this section, we further explore these results from Sect.~\ref{sec:galaxy_mah} and ~\ref{sec:stellar_pops} and compare them with several related works in literature. 
We examine the physical drivers behind the stellar populations and mass-assembly histories that we observe today in the bulges and discs of spiral galaxies, and how they fit into the broad picture of galaxy formation and evolution.

\subsection{Implications on galaxy stellar populations}
\label{subsec:stellar_pops_implications}

From Figures \ref{fig:bulge-disc-met-violin} and \ref{fig:bulge-disc-age-violin}, we find that our sample can mostly be split up into two (major) different stellar populations - one with bulges that are older and more metal-rich than discs, and one with bulges that are younger and more-metal rich than discs - which are both discussed in detail below. In this section, we discuss certain trends in terms of the total galaxy stellar mass (defined here as the simple sum of bulge and disc stellar mass for a galaxy), so that it fits in the context of the literature.

\paragraph{Case I: Galaxies with older and more metal-rich bulges.}
For the full spiral galaxy sample, our results show that this sample is dominated by bulges that are more metal-rich than their disc counterparts. On average, $\sim 68\%$ of the spiral galaxies contain older bulges than the discs, $\sim 79\%$ contain more metal-rich bulges than the discs, $\sim 51\%$ contain both older and more metal-rich bulges. This trend agrees with many studies in literature for S0s and spirals \citep{macarthur2009sp, sanchez-blazquez2011sfh, johnston2012fornax, johnston2014s0, sanchez-blazquez2014, goddard2017, breda&papaderos2018, barsanti2021sami, parikh2021sp, johnston2022buddi2}. However, it must be noted that most of these studies compare the stellar populations of bulge and disc-dominated regions of the galaxy measured in terms of radii rather than the bulge-disc decomposition approach we take in this work. 

Our stellar age and metallicity results also point to a dependence on total galaxy stellar mass. By dividing the sample into 4 mass bins, we find that $\sim 53\%$ of the galaxies in the lowest mass range  $\mathrm{M_* < 10^{9} M_{\odot}}$, have older bulges than discs, although this bin only has 19 objects (low-number statistics). For $\mathrm{10^9 M_{\odot} < M_* < 10^{10} M_{\odot}}$, this population makes up $\sim 40\%$ (the younger bulges dominating this mass range are discussed later in this section). In the mass bin  $\mathrm{10^{10} M_{\odot} < M_* < 10^{11} M_{\odot}}$, the fraction of older and more metal-rich bulges increases to $74\%$, and for galaxies with  $\mathrm{M_* > 10^{11} M_{\odot}}$, this further reaches a maximum at $96 \%$. 

A similar (total) galaxy stellar mass binning to investigate stellar metallicities shows that out of the galaxies with $\mathrm{M_* < 10^{9} M_{\odot}}$, $\sim 68\%$ host bulges that are more metal-rich than discs. This fraction increases with galaxy stellar mass, with $\sim 72\%$ in the mass bin $\mathrm{10^9 M_{\odot} < M_* < 10^{10} M_{\odot}}$, $\sim 82\%$ in $\mathrm{10^{10} M_{\odot} < M_* < 10^{11} M_{\odot}}$, and $\sim 87\%$ for galaxies with $\mathrm{M_* > 10^{11} M_{\odot}}$.

The relatively high metallicity in the majority of the bulges can be explained on the basis of the depth of the gravitational potential well \citep{barone2020grav, lah2023}. The bulges, having a deeper potential well, can retain their metals more strongly than the discs, whose shallower potential makes it easier to lose their metals to the intergalactic medium. The metallicity difference between the bulge and the disc $\mathrm{\Delta [M/H]_{bulge-disc}}$ is fairly constant within their error bars, for galaxies of stellar masses over $\mathrm{10^{10} M_{\odot}}$, but show no significant trend. Below this mass, our sample has relatively fewer galaxies, and shows a wider range in $\mathrm{\Delta [M/H]_{bulge-disc}}$. This result also agrees with \citet{dominguez-sanchez2020}, where they find a lack of metallicity trends for galaxies over $\mathrm{3\times10^{10} M_{\odot}}$. However, contrary to us, they find strong metallicity gradients for low-mass galaxies in their sample. For the very early type spirals (Sa and Sab), these stellar populations could potentially signal a phase of their evolution as they transform to S0 galaxies. S0s are considered to be an endpoint in the evolution of spiral galaxies once the gas has been used up or stripped away and the star formation truncated. However, in order to better understand this transformation, we must first understand the properties of the bulges and discs in the progenitor spirals, as we explore in this study. An inside-out quenching scenario has been proposed in \citet{mendel2013quench}, \citet{tacchella2015quench}, and \citet{barsanti2021sami}, where star formation is shut off first in the bulge as a result of AGN feedback or halo quenching through the mechanisms described in Sect.~\ref{sec:intro}. The quenching phase then propagates outwards from the bulge across the disc, which eventually leads to a morphologically transformed galaxy - an S0 - with a bulge older than the disc.

Our sample of spiral galaxies shows older stellar ages for bulges than for the discs, for galaxies of stellar masses above $\mathrm{10^{10} M_{\odot}}$. This agrees with results obtained in previous studies \citep{sanchez-blazquez2014, goddard2017, fraser2018manga, pak2021califa, johnston2022buddi2}. This is often explained by the fact that discs have ongoing star-formation fuelled by accreting gas, while star-formation has long ceased in the bulge. Similar to the metallicity differences, for masses below $\mathrm{10^{10} M_{\odot}}$, there are fewer galaxies but those show a wider range of stellar age differences between the bulge and the disc. 

\paragraph{Case II: Galaxies with younger and more metal-rich bulges.}
As we noted in Sect.~\ref{sec:stellar_pops}, although the older and more metal-rich bulges dominate our sample, there is a significant fraction of galaxies that host a different stellar population - with younger and more metal-rich stars in the bulge compared to the disc, forming $\sim 28\%$ of the sample. In our sample, this fraction appears to depend on morphology, and increases steadily with each type - with $\sim 5\%$ in Sa spirals and ending with $\sim 57\%$ in Scd spirals. This coincides well with the findings in \citet{barsanti2021sami}, where they study the stellar populations of bulges and discs of S0 galaxies. \citet{bedregal2011} find that the presence of a galaxy population exhibiting a bulge that is younger and more metal-rich compared to its disc suggests that the bulge has experienced more recent star formation, fuelled by enriched material, in contrast to the disc where star formation activity has ceased. Along those lines,
\citet{johnston2014s0} suggest that the bulge in similar S0 galaxies has been fuelled by enriched gas, inducing more recent star formation activity there. They propose that while the disc is being quenched, the gas collapses towards the inner regions of the galaxy, and incites a final episode of star formation, which biases their mean luminosity-weighted stellar populations in the bulge towards younger ages. Another possibility is presented in \citet{coelho&gadotti2011}, where they find by comparing the ages and metallicities of bulges hosted by barred and non-barred galaxies, that bars play a significant role in funnelling gas from the disc along the bar towards the central bulge. This induces a new episode of star formation there, pushing the observed stellar populations towards younger ages. However, given the low spatial resolution and field of view in MaNGA, modelling reliably a third component representing the bar would prove difficult, especially in an automated fashion as for BUDDI-MaNGA. Therefore, we continue with the simple two-component model for this work. For further investigation on the properties of bars as an independent component of the galaxy, and the effect they could have on the relative stellar populations of bulges and discs would be better studied using other IFUs with a better spatial resolution such as MUSE. However, this is beyond the scope of the BUDDI-MaNGA project.

Another possible reason for this trend has been discussed and debated extensively in the literature as the presence of `pseudo-bulges' in these galaxies. These disc-like pseudo-bulges are often the result of a steady and secular evolution of the disc over sufficiently long timescales with star forming episodes up to recent times, leading to the formation of a central bulge (although only named such because this component is more centrally concentrated than the disc) that mimics the properties of the precursory disc. These pseudo-bulges have been found to contain sufficient amounts of cold gas needed to build up its stellar mass through continued star formation, fuelling their internal evolution \citep{gadotti&anjos2001bulges, kormendy&fisher2008, peletier2008bulges, erwin2015bulges, kormendy2016}.
Hence, these bulges would be expected to be relatively younger on average or have similar ages to their corresponding discs. In our work, we do not explicitly classify the bulges as classical or pseudo since most criteria in literature involves an arbitrary selection cut in S\'ersic index, which can lead to significant mixing considering not all literature agree on the exact selection cut. Furthermore, \citet{haeussler2022galfitm} find that while modelling galaxies with \textsc{Galfit} and \textsc{GalfitM}, the S\'ersic index is the hardest structural parameter to accurately recover, and hence cannot be reliably used to separate different bulge types. We concede that bulges do not come in distinct classes, but rather exist on a spectrum with the possibility of more than one kind of bulge existing in a galaxy \citep{erwin2015bulges}. Furthermore, it is plausible that a fraction of the galaxies in our sample also host such a composite bulge system containing different stellar populations, although this is much harder to disentangle. 

In that vein, \citet{breda&papaderos2018} determine the driving mechanisms behind the formation of bulges in late-type galaxies, and subsequently discuss the validity of the distinctness of classical and pseudobulges. Their sample consisted of 135 late-type galaxies in the CALIFA survey, on which 2D spectral modelling had been performed, alongside surface photometry of their corresponding SDSS images. In accordance to the results presented in this work, they find that the most massive bulges host the oldest and most metal-rich stellar populations. The mass-weighted stellar ages and metallicities of the bulges and discs showed a homologous increasing trend with the galaxy stellar mass. For the lowest-mass galaxies, the bulges and the discs exhibited nearly indistinguishable stellar population properties. In the higher mass galaxies, the bulges were clearly older than the discs, having formed $\sim2-3$ Gyr earlier (readers can refer to their Fig. 7), consistent with our results. On observing the difference between the stellar age of the bulges and the discs as a function of galaxy stellar mass, they found a strong positive trend. A weaker but significant positive trend was observed for the metallicity difference as a function of galaxy mass. They also find no evidence of an age bimodality that would confirm the existence of two distinct bulge populations; they imply instead that bulges and discs evolve parallel to each other. For a more exhaustive list of the various stellar populations gradients and trends observed in various studies, we refer the reader to the appendix in \citet{lah2023}.

\subsection{Implications on galaxy assembly: downsizing and growth from the inside-out}
\label{subsec:assembly_mode_implications}

We have studied the mass-assembly histories of bulges and discs of our spiral galaxy sample independently as a function of their respective component stellar masses and their morphologies (type). 
The bulges, especially in early-type spirals show a prominent downsizing trend, with the stellar populations in high-mass bulges having assembled notably earlier than those in their low-mass counterparts. In discs, the downsizing effect is diluted but still exists in the earlier types, and breaks down in the later types (Fig. \ref{fig:sfh_individual} and Fig. \ref{fig:sfh_mass_dependence}). This downsizing trend we observe in our sample complies with the results of previous studies in a qualitative sense, although a direct quantitative comparison would not be possible owing to the differences in methodologies and galaxy samples employed in each study. 
 
For all the spirals in our sample of (total) galaxy stellar mass $M_{*}<\mathrm{10^9 M_{\odot}}$, the discs appear to have assembled half their stellar masses approximately $0.12^{+7.3}_{-6.5}$ Gyr  before the bulges did. Similarly, galaxies in the mass range $\mathrm{10^9 M_{\odot} < M_* < 10^{10} M_{\odot}}$ on average have assembled half the mass in their discs $0.89^{+4.4}_{-5.4}$ Gyr before their bulges. These low mass galaxies are the most diverse in terms of their MAHs and therefore the median is only a rough estimate of the time window between bulge and disc mass assembly, but nevertheless points to an outside-in assembly scenario on average. However, considering the high uncertainties on recovering stellar ages beyond 8 Gyr \citep{ibarra-medel2016, ibarra-medel2019}, some of these galaxies might inherently have gone through an inside-out assembly or simultaneous bulge and disc assembly, which is also suggested from the diversity of MAHs in the components of these low mass galaxies.  
Higher mass galaxies within the range $\mathrm{10^{10} M_{\odot} < M_* < 10^{11} M_{\odot}}$ show a clear inside-out assembly mode with a half-mass formation time difference of $2.02^{+6.6}_{-1.8}$ Gyr. The inside-out assembly is the most pronounced in the galaxies of the highest stellar masses: $\mathrm{M_* > 10^{11} M_{\odot}}$ with the discs having assembled half their masses by about $4.5^{+9}_{-0.9}$ Gyr on average after the bulges did. As we can see from the uncertainties denoted by the 16th and 84th percentiles of the timescale distribution, these median timescale estimates are rough and do not represent the diversity of MAHs in their entirety. It must be noted at this point that while these MAHs trace the time when the stellar populations have assembled, they do not give us any information on their origin, that is, whether they are stars born in situ in the gravitational potential of the galaxy, or if they are ex situ stars accreted in mergers. 

From Figure~\ref{fig:sfh_ttype_dependence}, our results indicate the MAHs of bulges and discs depend moreover on the morphological types. The later-type spiral galaxies show no clear distinction between the disc and bulge star formation histories, with a much lesser dependence on their stellar masses as compared to the early-type spirals. Additionally, these galaxies lie in the low-mass regime of our sample and could imply that their bulges and discs have varied mass assembly histories with either inside-out or outside-in modes. A single assembly mode might dominate consistently throughout the lifetime of a galaxy, or different modes might dominate at different epochs which is also implied by some bulge and disc MAHs transitioning after some Gyr. Some of these bulges and discs could have formed together from the same material, over similar extended timescales, with several bursts of star formation throughout their lifetime. However, there is also the possibility that these late-type galaxies are in fact bulge-less and our assigning two components consistently for all galaxies in the sample results in similar mass-assembly histories for these in particular. Although a quantitative flagging of galaxies that are best represented by a single-component model or a two-component model is beyond the scope of this paper, we plan to address this in future work. 

The inferences from our study regarding the formation pathways of galaxies match those from \citet{perez2013}, where their sample of galaxies were found to have grown their stellar mass from the inside-out. With 105 objects from the CALIFA survey, they employ the fossil record method to reconstruct spatial mass growth histories across cosmic time. The spectra were extracted in four spatial galaxy regions from the centre to the outskirts, defined as a function of the half-light radius $R_{50}$. This pivotal study was the first to present evidence of galaxy downsizing using integral field spectroscopy (IFS). The mass growth histories were quantified by the time it takes for 80\% of the stellar mass to have been built up. They found that the outer-most regions of less massive galaxies had reached this mass fraction as recent as about 1 Gyr ago, while the more massive ones had reached the same fraction almost 5 Gyr ago - clearly indicating that the downsizing phenomenon depends on the galaxy stellar mass. Their results support the inside-out formation theory, where the stars had assembled in the inner regions much earlier than in the outer regions. However, at galaxy stellar masses $M_*<10^{9.58}M_{\odot}$, they observe a transition to outside-in formation similar to that observed in dwarf galaxies. In the framework of formation pathways, they propose that the high-mass galaxies had grown their stellar masses in the inner regions first as a result of a merger event about $5-9$ Gyr ago, while the lowest-mass galaxies had instead undergone a steady secular evolution, a result that is supported by our findings as well.

Our results also qualitatively confirm the findings in \citet{ibarra-medel2016}, where the primary focus was of a similar nature to understand how galaxies assembled their stellar masses, using SDSS-MaNGA DR13. With the spatial information available, they studied the mass growth histories (MGHs) at different radial regions up to 1.5$R_e$, as a function of galaxy stellar mass and morphology. Moreover, they also probed the assembly as a function of other galaxy properties such as specific star formation rate and colour. The difference in methodology is that we use distinct bulge-disc decomposition cleanly separating out the spectra from each component, while they separate the galaxies into `inner' and `outer' regions in relation to the half-light radius. Nevertheless, they find that the more massive a galaxy is, the earlier is its assembly (quantified by the 50\%, 70\%, and 90\% mass formation time, which they define as the LBT difference between the inner and outer regions when the MGH reaches these percentages of stellar mass). They also find a large diversity and scatter in the assembly histories of dwarf and low-mass galaxies, with more extended and episodic star formation.  In terms of the assembly mode, the innermost regions of most of their galaxies had assembled their stellar populations before the outermost regions, pointing to the inside-out formation mode as in ours. Another difference between our analyses is that they study the dependence of MGHs on the galaxy stellar mass, while we note the dependence on the component stellar mass (how the bulges assemble as a function of bulge stellar mass, and similarly for the discs). 

Another similar SDSS-MaNGA study is described in \citet{peterken2020manga}, where they built the spatially resolved star-formation histories using the full-spectrum fitting software \textsc{STARLIGHT} \citep{cidfernandes2005starlight}. Their approach involved creating time-slice maps, which allowed them to look at regions where stellar populations of any given stellar age are present. For their sample of 795 spiral galaxies, they again find clear evidence for inside-out galaxy growth, with a negative age gradient. The outskirts of the galaxies are consistently younger than the central regions in 80\% of the sample for galaxies of stellar mass $M_* > 10^{10.22} M_{\odot}$. They complement this result by defining the formation time as when 95\% of the total galaxy mass was assembled in each region of the time-slice, which decreases radially from the centre. This mode of galaxy growth becomes predominant at the highest masses, in line with our generalised results.

\citet{breda&papaderos2018} (described earlier in Sect. \ref{subsec:stellar_pops_implications}), in accordance to the results presented in this work, find that the most massive bulges are hosted by massive late-type galaxies ($M_*\gtrapprox10^{10.7}M_{\odot}$). They note that the timescale of mass assembly is shorter for the most massive bulges compared to the less massive ones, which assemble their masses over longer timescales. Furthermore, the high-mass bulges are found to be on average older than their low-mass counterparts by $\sim4$ Gyr, where the galaxy component downsizing is clear. They also find evidence of the oldest bulges being hosted by the oldest discs. The authors suggest that bulge growth is driven by a combination of faster and earlier processes namely a monolithic collapse, and slower secular processes. In this scenario, the bulges and discs are expected to grow and evolve in an interwoven fashion, with the bulge assembling out of the parent or hosting disc. The bulge might then continue to grow through processes such as inward stellar migration. This interpretation also implies that the Sérsic index of the bulge increases with galaxy stellar mass, which is mirrored in our study (Appendix~\ref{appendix4}).  We therefore observe that there are clear similarities in the results of the stellar population analysis and formation mechanisms of bulges observed in this study and the present one.

Our results also tie in with the findings in \citet{mcdermid2015atlas}, where they present a similar analysis on the ETGs in the IFU survey $\mathrm{ATLAS^{3D}}$ \citep{cappellari2011atlas}. The goal of their work was not to study the bulge and disc components or the inner and outer regions of the galaxies, but rather their star formation histories and stellar populations as a whole, as a function of dynamical mass and environment. Nevertheless, they find that the SFH depends on the present day dynamical mass of the galaxy for their sample, and that global galaxy downsizing prevails - the more massive galaxies had been formed much earlier than the less massive galaxies. Furthermore, they also measure the mean SFR and SFR density which allowed them to conclude that in ETGs the less massive galaxies display more extended star formation histories, while in the most massive ones, the star formation rapidly falls off leading to quiescence. They quantify the stellar mass assembly by their half-mass formation time and find that their most massive galaxies ($\mathrm{10^{11.5} M_{\odot} < M_{dyn} < 10^{12} M_{\odot}}$) have formed their stellar populations by 2 Gyr after the Big Bang. On the other hand, the least massive galaxies in their sample ($\mathrm{10^{9.5} M_{\odot} < M_{dyn} < 10^{10} M_{\odot}}$) only finish assembling 50\% of their masses by 8 Gyr after the Big Bang. 

In line with ETGs, \citet{johnston2022buddi2} was a precursor to this work, which used \textsc{BUDDI} for bulge-disc decomposition of S0 galaxies in SDSS-MaNGA DR15, and whose procedure of estimating mass-weighted stellar populations and mass assembly histories using \textsc{pPXF} we have followed here. Consistent with the literature listed above and the current work, they find primarily an inside-out formation scenario for high-mass S0s with $\mathrm{M_* > 10^{10} M_{\odot}}$. The bulges assemble their stellar populations first rapidly, and the disc follows with a diverse range of star formation histories extending up to recent times. In the lower-mass galaxies, they find no evidence of a consistent mass assembly mode, owing to the diversity in SFH. However, their results support downsizing in bulges with respect to the total galaxy stellar mass, while the discs show no evidence of this, atleast in the S0 galaxies they examined. We also note that their sample size is much smaller, with only 78 galaxies, so these differing inferences could be real in that spiral galaxies and S0s inherently do not have the same assembly pathways, or instead just be the result of a lack of a statistically significant sample. While our results here support downsizing in both bulges and discs, albeit at a weaker level in the discs, they are studied in terms of the component masses and not the total galaxy stellar mass as they do. 

More recently, a study based purely on imaging data from GAMA, \citep{bellstedt2023profuse} found results that are in accordance with ours in this work. In their approach, they used \textsc{ProFuse}, which is another spectro-photometric technique (where the spectral aspect encompasses the UV to the far infrared) that models the 2D surface brightness profile of galaxy components in multiple wavelengths simultaneously and derives their stellar populations, by combining both structural decomposition and SED fitting. This technique is fairly similar in concept to \textsc{BUDDI}. With \textsc{ProFuse}, they model a sample of 7\,000 galaxies in GAMA to study the cosmic star formation history (CSFH) of bulges and discs. They found that the stars in the discs had formed relatively recently within the last 8 Gyr. The bulges were found to host the oldest stellar populations, where a majority of the stellar mass had been formed 11.8 Gyr ago. The CSFH that were reconstructed with \textsc{ProFuse} (simultaneous profile fitting and SED fitting) was also consistent with that derived from SED fitting with \textsc{ProSpect} (each step done consequently). This agreement among the different techniques (including \textsc{BUDDI}) in deriving the general stellar populations of bulges and discs emphasises the reliability of all these techniques and our understanding of when the different components assembled their stellar masses.

To recap, previous IFS-based studies on inside-out galaxy assembly, and galaxy and component downsizing, and the dependence of these phenomena on stellar mass \citep{perez2013, breda&papaderos2018, johnston2022buddi2} is consistent with the insights from the present study.

\section{Conclusions}
\label{sec:conclusions}

In summary, we have performed bulge-disc decomposition of integral-field spectroscopic observations and extracted the clean spectra of both components for 1452 galaxies in SDSS-MaNGA DR17. Following this, for the spiral galaxy sample of 968 objects, we derived the mean mass-weighted stellar populations hosted by each component through full spectral fitting using \textsc{pPXF}. With the mass weights obtained during the spectral fitting, we have traced the mass-assembly histories of the bulges and discs through cosmic time, and investigated their dependence on component mass and morphology. Our main findings are listed below:

\begin{itemize}
    \item Through the fossil-record analysis, we find evidence of a clear global downsizing trend in terms of the component stellar masses, for spiral galaxies of types Sa-Sc. The effect is more prominent in bulges and is diluted in the discs. For the late-type spirals Scd-Sd, downsizing breaks down and the mass assembly becomes more random on average.

    \item We find that bulges on average assemble their stellar masses rapidly, possibly in a single episode of star formation, during the very early stages of their lifetime. The discs however take longer to assemble their masses with more delayed and extended star formation histories. This effect is also entangled with the morphology dependence, where with increasing types, the bulges also begin to show a larger fraction of extended mass assembly. By the latest spiral types Scd-Sd, almost all galaxies show similar delayed mass-assembly histories in their bulges and discs.

    \item The individual mass-assembly histories of the bulges and discs were quantified by the half-mass formation time, in which 50\% of the stellar mass has been built up by the respective component. The MAHs had a clear dependence on both stellar masses and morphologies. From Sa-Sc types, the main mode of mass growth is from the inside out, and by separating into four total galaxy stellar mass bins, we find this mode to be strongest in the most massive galaxies. For the late-type spirals Scd-Sd, it becomes harder to identify a specific formation mode and they rather show a large diversity in their MAHs, suggesting different epochs where different assembly modes are prevalent. 

    \item The above-mentioned result is also complemented by stellar population analysis. The majority of spiral galaxies contain bulges that are older and more metal-rich than the discs. A smaller but significant fraction contain bulges that are younger and more metal-rich than their discs, and this trend becomes stronger with increasing morphological types.

\end{itemize}    

In conclusion, this work is the third paper within the BUDDI-MaNGA project, where we present the results from the newest version of the fits with SDSS-MaNGA DR17. The galaxies were modelled with a S\'ersic + Exponential profile for the bulge and disc, respectively, with the added advantage of incorporating masks for neighbouring foreground stars to improve the fits from the previous version described in Paper I and Paper II. This forms the largest sample to date of 1452 clean bulge and disc spectra derived from IFU datacubes, allowing detailed statistical analyses of their stellar populations. Our results suggest that spiral galaxies in general assemble their stellar mass through the inside-out pathway, but it is not the sole mode of assembly. Furthermore, our results also show two major stellar populations, with bulges being older or younger and more metal-rich than their discs, again suggesting more than one pathway of stellar mass assembly in spirals. This spectroscopic study of spirals is one of several within BUDDI-MaNGA, the goal of which is to better understand the formation and evolution of galaxy bulges and discs across a range of morphologies and stellar masses, and how they fit into the broader picture of galaxy evolution.

\begin{acknowledgements}

We thank the anonymous referee for their careful, constructive report and suggestions to improve the paper. KJ acknowledges financial support from ANID Doctorado Nacional
2021 project number 21211770. KJ thanks the LSST-DA Data Science Fellowship Program, which is funded by LSST-DA, the Brinson Foundation, and the Moore Foundation; their participation in the program has benefited this work. EJJ acknowledges support from FONDECYT Iniciaci\'on en investigaci\'on 2020 Project 11200263 and the ANID BASAL project FB210003. KJ and EJJ acknowledge financial support of Millenium Nucleus ERIS NCN2021\_017. KJ thanks Payel Das, Alfonso Aragón-Salamanca, Shravya Shenoy, and Robert Yates for useful discussions.

Funding for the Sloan Digital Sky Survey IV has been provided by the Alfred P. Sloan Foundation, the U.S. 
Department of Energy Office of Science, and the Participating Institutions. 

SDSS-IV acknowledges support and resources from the Center for High Performance Computing  at the University of Utah. The SDSS website is www.sdss.org.

SDSS-IV is managed by the Astrophysical Research Consortium for the Participating Institutions of the SDSS Collaboration including the Brazilian Participation Group, the Carnegie Institution for Science, Carnegie Mellon University, Center for Astrophysics | Harvard \& Smithsonian, the Chilean Participation Group, the French Participation Group, Instituto de Astrof\'isica de Canarias, The Johns Hopkins University, Kavli Institute for the Physics and Mathematics of the Universe (IPMU) / University of Tokyo, the Korean Participation Group, Lawrence Berkeley National Laboratory, Leibniz Institut f\"ur Astrophysik Potsdam (AIP),  Max-Planck-Institut f\"ur Astronomie (MPIA Heidelberg), Max-Planck-Institut f\"ur Astrophysik (MPA Garching), Max-Planck-Institut f\"ur Extraterrestrische Physik (MPE), National Astronomical Observatories of China, New Mexico State University, New York University, University of Notre Dame, Observat\'ario Nacional / MCTI, The Ohio State University, Pennsylvania State University, Shanghai Astronomical Observatory, United Kingdom Participation Group, Universidad Nacional Aut\'onoma de M\'exico, University of Arizona, University of Colorado Boulder, University of Oxford, University of Portsmouth, University of Utah, University of Virginia, University of Washington, University of Wisconsin, Vanderbilt University, and Yale University.

\end{acknowledgements}

\bibliographystyle{aa} 
\bibliography{papers} 

\begin{appendix} 

\section{SDSS photometry with BAGPIPES - Stellar mass dependence of e-folding time $\tau$ }
\label{appendix1}

In Sect. ~\ref{subsec:mah_mass_dependence}, the MAH built from spectral fitting show a clear dependence on the component stellar mass. In the context of our photometric analysis, we again find that the more massive the component is, the quicker its assembly. The bulges and discs with the highest masses had assembled their stellar masses in a shorter timescale. The bulges also showed a higher fraction of rapid assembly compared to the discs, apart from the observed mass dependence.

\begin{figure}[h]
    \centering
    \includegraphics[width=\columnwidth]{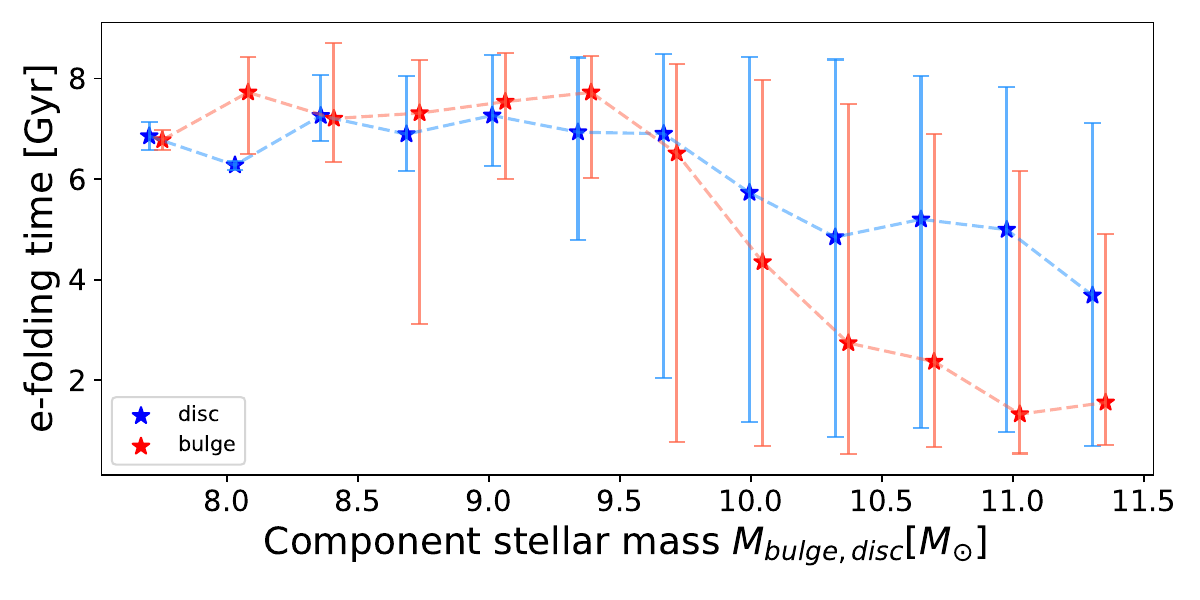}
    
    \includegraphics[width=\columnwidth]{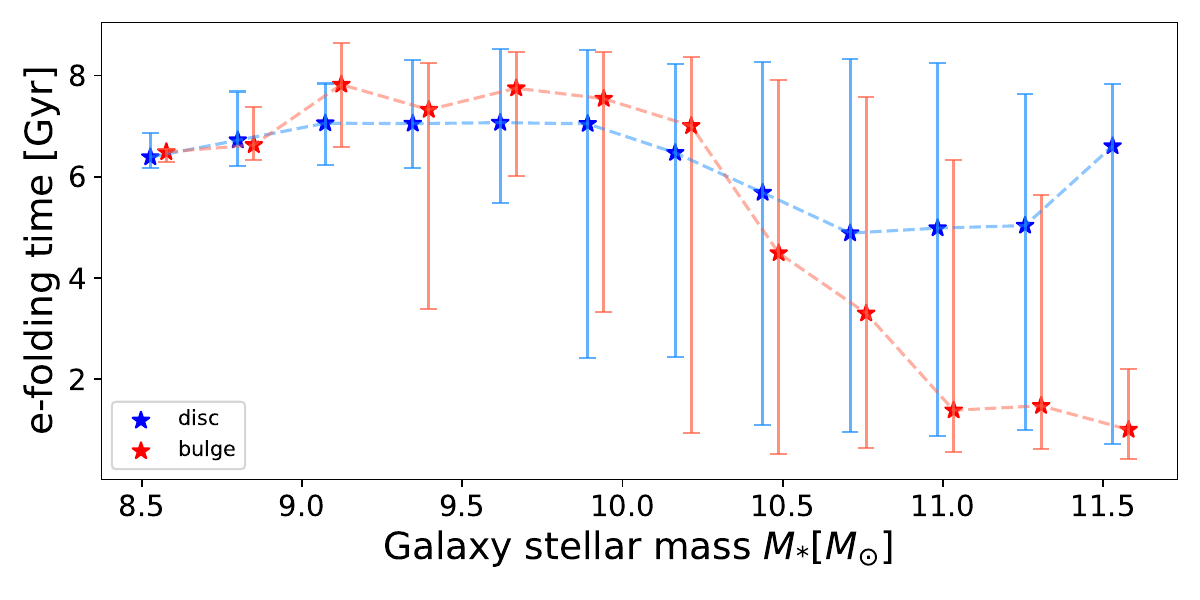}    
    \caption{Dependence of e-folding time as a function of stellar mass. Upper panel: Variation of e-folding times of bulges $\tau_{bulge}$ (red) and discs $\tau_{disc}$ (blue) as a function of their corresponding component stellar masses. The stars represent the median value in each mass bin, and the error bars show the 16th-84th percentile range in the spread of values in each bin. Lower panel: Similar to above but with the total galaxy stellar mass along the x-axis. In both panels, the $\tau_{bulge}$ was shifted by 0.05 dex for clarity in visualisation.}
    \label{fig:etau}
\end{figure}

In this section, we compare these global results from spectra to those we obtain from SED fitting of the corresponding SDSS photometry of the galaxies in the $ugriz$ bands using BAGPIPES. In the setup for the SED fitting, we assume a parametric star-formation history that is exponentially declining ($\tau$ model), where $\tau$ is the e-folding time, which is the timescale needed for the star formation rate (SFR) to decline by a factor of $e$. This parameter essentially gives us information about star formation timescales in the galaxy bulges and discs - the shorter the e-folding time, the shorter is the star formation period. On the other hand, a longer e-folding time implies a more extended star formation period. 

From the upper panel of Figure \ref{fig:etau}, the median $\tau$ of galaxy bulges $<10^{10} M_{\odot}$ show relatively high values between 6 and 8 Gyr, with no trend within the bins of this bulge mass range. These high e-folding times indicate slower star formation timescales over an extended period. After this, $\tau_{bulge}$ drops steeply with increasing stellar mass. This indicates that these higher mass bulges would have experienced a burst of star formation where a significant fraction of its stars formed over a short period, which becomes shorter with increasing bulge mass (dropping from $\sim 6$ Gyr to $\sim 1$ Gyr in the bins $>10^{10} M_{\odot}$). The discs show a very similar trend with the disc stellar mass, in the low-mass regime of $<10^{10} M_{\odot}$, where the e-folding times of both the bulges and discs appear to almost overlap with each other. At these lower masses defined earlier, the long and extended star formation dominates. At higher masses, the e-folding times of the disc again declines with increasing disc mass, but the trend is shallower compared to the bulges and slowly drop by 1 - 2 Gyr. We note that the uncertainties on the median $\tau$ are quite large, spanning almost 8 Gyr in some cases, implying a large diversity in the star formation timescales, when we observe the dependence purely on the stellar mass. Nevertheless, the median trends agree well with our results from spectral analysis in Sect.~\ref{subsec:mah_mass_dependence}. 

In the lower panel of Figure \ref{fig:etau}, we observe the same general trend of the e-folding times as a function of the total galaxy stellar mass instead of the component mass, with the decline in $\tau$ beginning at $\sim 10^{10.5} M_{\odot}$. The decline in the discs is shallower (with the upturn at $\sim11.5 M_{\odot}$ explained by low-number statistics of high mass discs in our sample). We associate the shallow drop in $\tau_{disc}$ with total mass to the significant diversity in star formation histories in discs. We assume an exponentially declining SFH throughout for both bulges and discs to support an automated fitting process, but we acknowledge that some of the discs might need additional starburst models superimposed on the standard $\tau$ model.  

\section{Bulge Sérsic index as a function of galaxy stellar mass}
\label{appendix4}

\begin{figure}[h]
    \centering
    \includegraphics[width=\columnwidth]{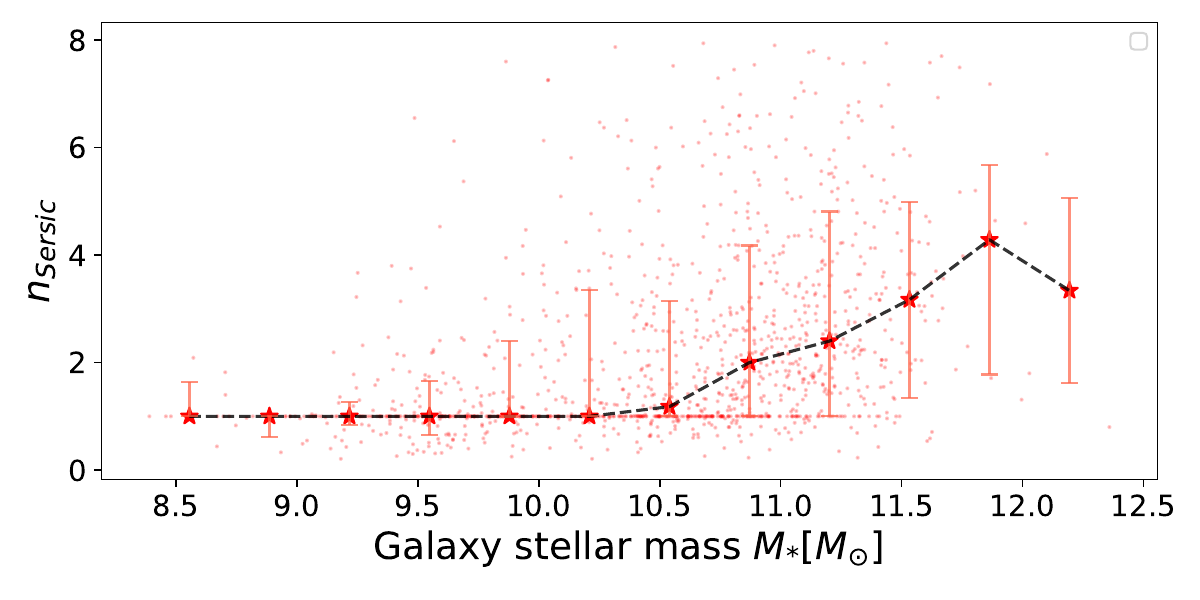}   
    \caption{Bulge Sérsic index as a function of the total galaxy stellar mass $M_*$. The red points show the individual galaxies, while the red stars show the binned median, with the error bars marking the 16th-84th percentile range. For statistically significant mass bins of $M_* > 10^{10} M_{\odot}$, the Sérsic index of the bulge increases with increasing galaxy mass, which arises from the bulge assembling from the discs. The timescale of assembly is expected to be shorter for the more massive galaxies, and longer for the less massive ones, which is consistent with our results on downsizing seen in low-redshift spiral galaxies.}
    \label{fig:sersic_mass}
\end{figure}

\newpage

\section{Stellar population properties of bulges and discs}
\label{appendix2}

\begin{figure}
    \centering
    \includegraphics[width=0.9\columnwidth]{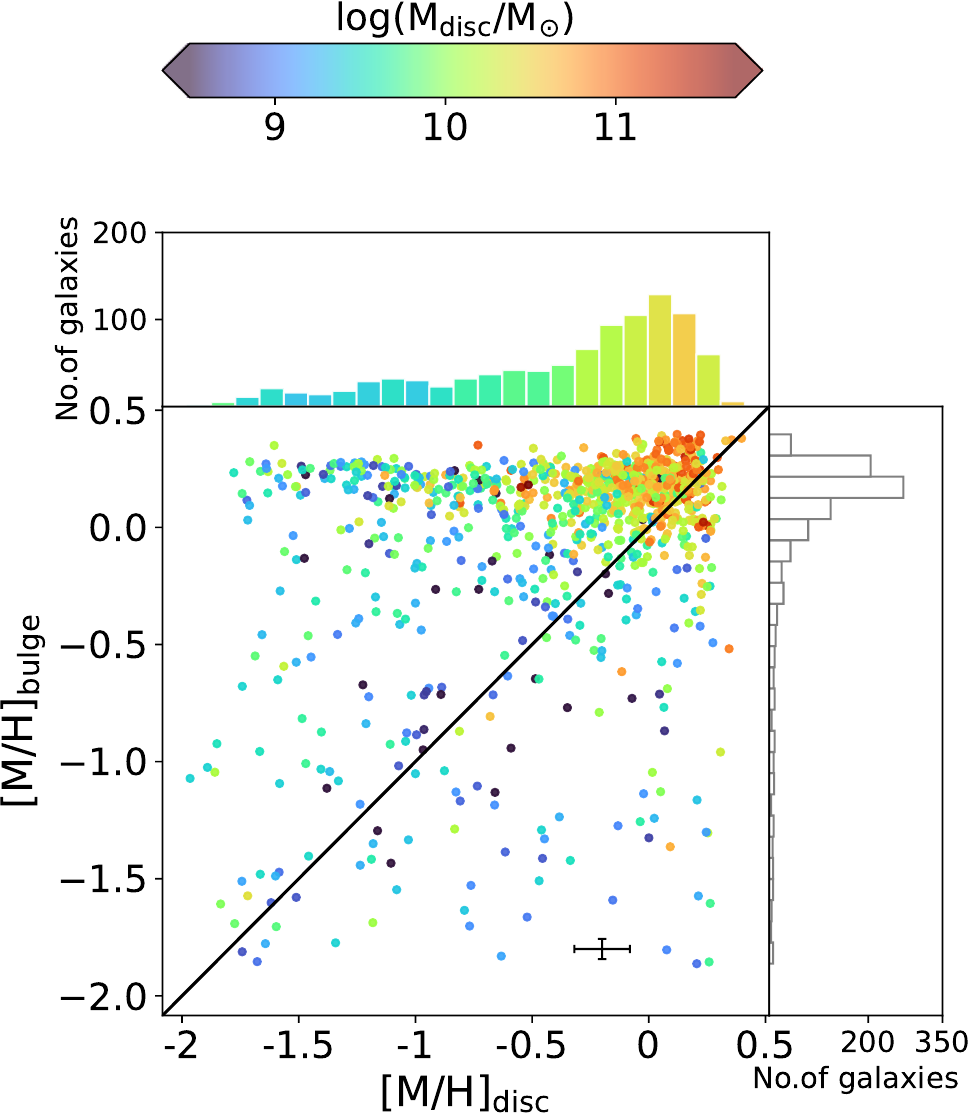}
    
    \includegraphics[width=0.9\columnwidth]{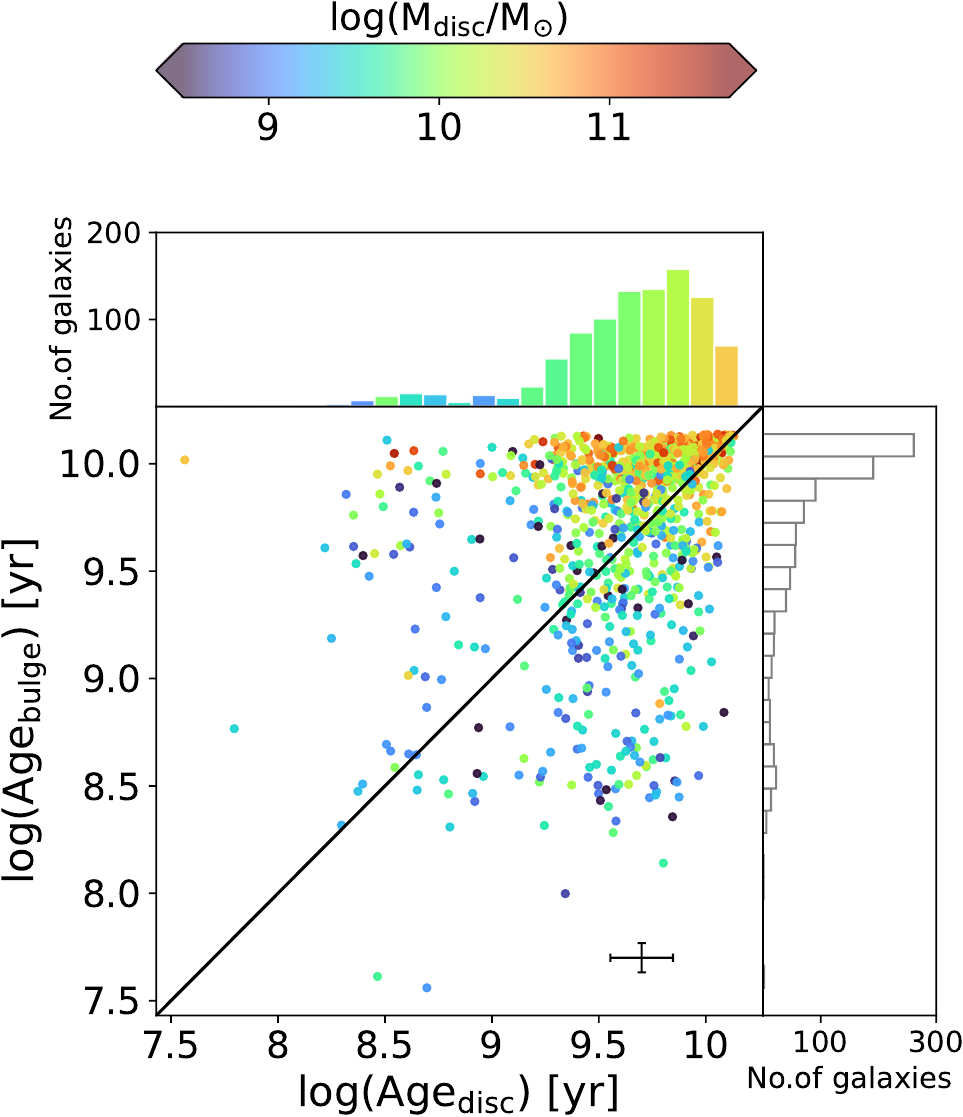}    
    \caption{Stellar population properties of bulges and discs and their corresponding distributions. Upper panel: Comparison of mean mass-weighted stellar metallicities of the bulges and the discs (with their uncertainties), colour-coded by disc stellar masses. The distributions in bulge (disc) metallicities are shown in the right (upper) joint histograms. Each bin in the metallicity histogram of the bulges is colour-coded by the mean disc mass in that bin. Lower panel: Similar to the upper panel but for stellar ages. The black diagonal line marks the 1:1 correlation.}
    \label{fig:met_age_trends_disc}
\end{figure}

In this section, we look at the dependence of the mass-weighted stellar populations on disc stellar mass in our spiral galaxy sample. In Figure \ref{fig:met_age_trends_disc} the upper panel compares the mean stellar metallicities of the bulge with the discs, as in Sect. \ref{sec:stellar_pops}, this time colour-coded by the disc masses instead. The lower panel compares the mean logarithmic stellar ages of the bulges and discs, colour-coded by the disc masses.  

From the upper panel, we find the same trends with respect to disc stellar mass as the bulge stellar mass in Sect. \ref{sec:stellar_pops}. The discs with the highest masses (red-orange points) again are clustered above the 1:1 line, implying that the majority of the spiral galaxies where the bulges are more metal-rich than the discs have both bulges and discs of high metallicities. With decreasing disc masses (towards the green-blue points on the left of the plot), we find galaxies with discs that have low metallicities. This general trend is clearly visible on the upper-joint histogram - starting with the massive high-metallicity discs (orange bins) on the right, steadily decreasing in both stellar mass and metallicity (blue bins) towards the left.

The lower panel shows a similar trend for stellar ages, wherein the most massive discs are also old (red-orange points). The histogram above shows a smooth trend from old and massive discs (orange bins) to young low-mass discs (blue bins).  

\section{Half-mass formation timescales of bulges and discs}
\label{appendix3}

\begin{figure*}
    \begin{center}
        \includegraphics[width=0.75\textwidth]{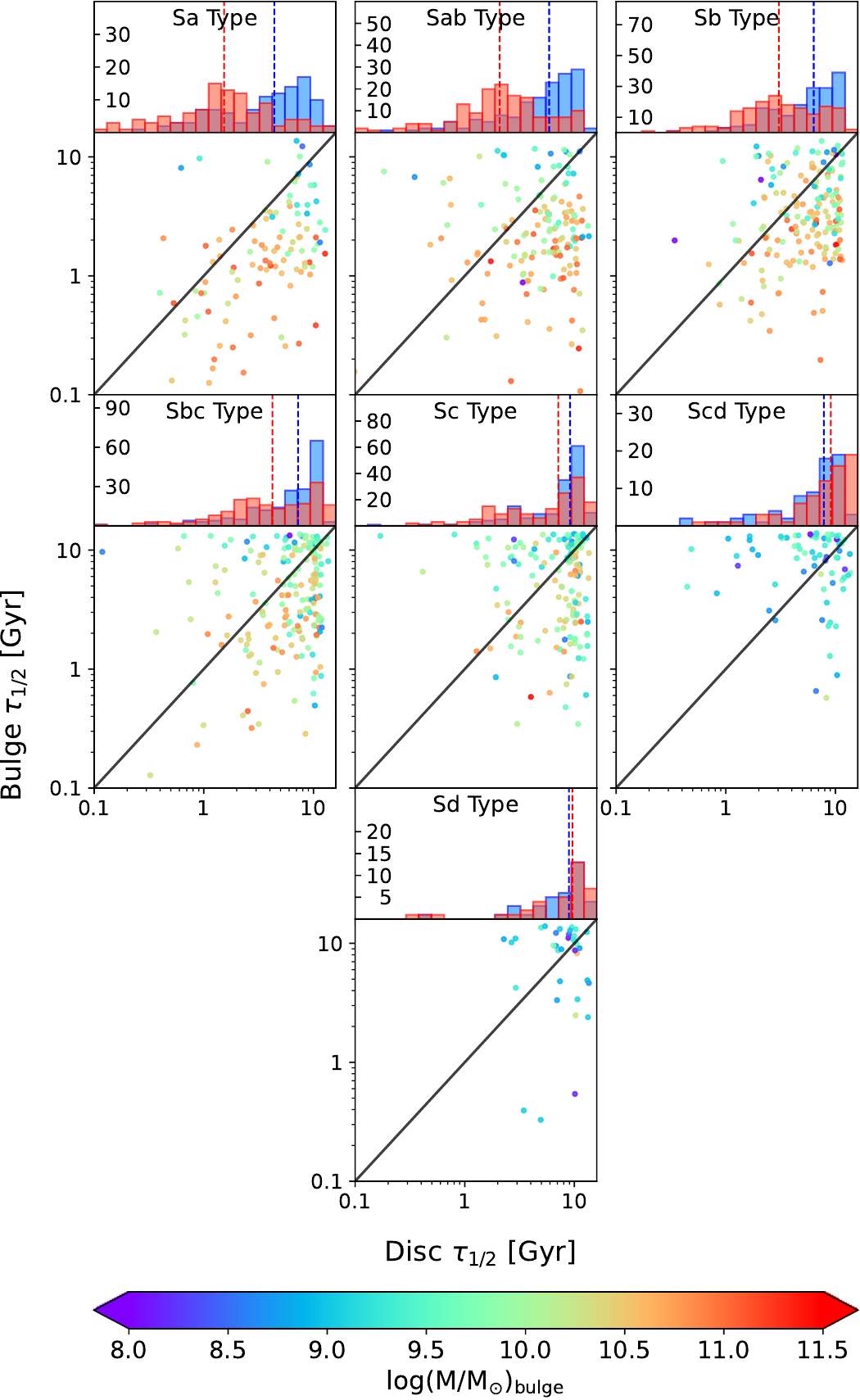} 
    \end{center}
    
    \caption{Half-mass formation timescales of the bulges and discs, colour-coded by bulge mass. The upper panels show the histogram of these formation timescales for the bulge and disc in red and blue respectively, with their median times in dashed lines.} 
    \label{fig:formation_timescale}

\end{figure*}

From Sect. \ref{subsubsec:half-mass}, we define the half-mass formation timescale ($\tau_{1/2}$), as the difference between the time when stellar mass assembly (star formation) first occurred in each component, and the time when 50\% of its current stellar mass had been assembled. The time of first star formation was taken to be the lookback time with the earliest non-zero weight measured by \textsc{pPXF} during the full-spectrum fitting. The results and inferences from Figure \ref{fig:formation_timescale} shown here are discussed in Sect. \ref{subsubsec:half-mass}.

\end{appendix}

\end{document}